# The Knowledge-Based Economy and the Triple Helix Model

Loet Leydesdorff

University of Amsterdam, Amsterdam School of Communications Research (ASCoR),
Kloveniersburgwal 48, 1012 CX Amsterdam, The Netherlands;
loet@leydesdorff.net, http://www.leydesdorff.net .

**Table of Contents**

## 1. Introduction – the metaphor of a "knowledge-based economy"

Few concepts introduced by evolutionary economists have been politically more successful than the metaphor of a 'knowledge-based economy.' For example, the European Summit of March 2000 in Lisbon was specifically held "to agree a new strategic goal for the Union in order to strengthen employment, economic reform and social cohesion as part of a knowledge-based economy" (European Commission, 2000).[1] Similarly and more recently, Barack Obama formulated in one of his campaign speeches: "[T]his long-term agenda […] will require us first and foremost to train and educate our workforce with the skills necessary to compete in a knowledge-based economy. We'll also need to place a greater emphasis on areas like science and technology that will define the workforce of the 21st century, and invest in the research and innovation necessary to create the jobs and industries of the future right here in America."[2]

Can such a large impact on the real economy be expected from something as poorly defined as the knowledge base of an economy? How can an economy be based on something as volatile as knowledge? In an introduction to a special issue on the topic, David and Foray (2002) cautioned that the terminology was coined recently, and noted

---

that "as such, it marks a break in the continuity with earlier periods, more a 'sea-change' than a sharp discontinuity" (*ibid*., p. 9). However, these authors also warned for confusion because the transformations can be analyzed at a number of different levels. They concluded that 'knowledge' and 'information' should be more carefully distinguished by analyzing the development of a knowledge-based economy in terms of codification processes (Cowan *et al*., 2000; Cowan and Foray, 1997).

Foray and Lundvall (1996) first introduced the concept of a 'knowledge-based economy' at a workshop of the Organization of Economic Cooperation and Development (OECD) in 1994 (OECD, 1996a). In that same workshop, Abramowitz and David (1996) suggested that *codified* knowledge should be made central to the analysis, and formulated as follows (at p. 35):

> Perhaps this single most salient characteristic of recent economic growth has been the secularly rising reliance upon *codified* knowledge as a basis for the organization and conduct of economic activities, including among the latter the purposive extension of the economically relevant knowledge base. While tacit knowledge continues to play critical roles, affecting individual and organizational competencies and the localization of scientific and technological advances, codification has been both the motive force and the favoured form taken by the expansion of the knowledge base.

Analytically, this focus on *codified* knowledge demarcated the new research program from the older concept of a 'knowledge economy' with its focus on knowledge workers and hence embodied knowledge (Cooke, 2002; Machlup, 1962; Penrose, 1959). Embodied and tacit knowledge is embedded in contexts (Bowker, 2005; Collins, 1974; Polanyi, 1961; Zuboff, 1988), while codified knowledge can be decontextualized, and therefore, among other things, traded on a market (Dasgupta & David, 1984). The metaphor of a knowledge-based economy appreciates the increased importance of organized R&D in shaping systems of innovation. The knowledge production function has become a *structural* characteristic of the modern economy (Schumpeter, 1939, 1943).

Whereas most economists have mainly been interested in the *effects* of codification on the economy more than in the processes of codification itself, and social scientists more in human knowledge than in knowledge as a social coordination mechanism, Daniel Bell formulated the possibility of this new research program already in 1973 as follows (at p. 20):

> Now, knowledge has of course been necessary in the functioning of any society. What is distinctive about the post-industrial society is the change in the character of knowledge itself. What has become decisive for the organization of decisions and the direction of change is the centrality of *theoretical* knowledge—the primacy of theory over empiricism and the codification of knowledge into abstract systems of symbols that, as in any axiomatic system, can be used to illustrate many different and varied areas of experience.

In other words, Bell postulated "a new fusion between science and innovation" (Bell, 1968, at p. 182). This fusion makes new institutional arrangements at the interfaces between systems possible (cf. Holzner & Marx, 1979; Whitley, 1984). The linear model of innovation in which basic research invents and industry applies with a one-directional arrow between them is replaced with an interactive and nonlinear one (Rosenberg, 1994, at p. 139; Godin, 2006a). In this context an improved understanding of the dynamic interplay between research, invention, innovation, and economic growth is required. In an age of changing practices of knowledge production and distribution, it is important to analyze how the communication of knowledge (e.g., discursive knowledge) and information relate and differ.

How do codifications within each sphere and knowledge transfer among institutional spheres interact? This project, in my opinion, requires an information-theoretical approach because of its focus on the relations between the production and distribution of both information and knowledge. While 'national systems of innovation' can be measured in terms of sectors and institutions, for example, by using national statistics, the ongoing globalization in a knowledge-based economy strongly suggests a theoretically guided research agenda at a supra-national level (Foray, 2004). Codification adds dynamic complexity to the multivariate complexity in the relations. Note that this project can be considered as 'infra-disciplinary'—that is, discipline and sector specific—unlike the 'trans-disciplinary' project of Mode-2 (Gibbons *et al.*, 1994; see below).

At the same OECD-workshop, Carter (1996) noted immediately that the measurement of a 'knowledge-based economy' poses serious problems. The OECD devoted considerable resources for developing indicators of 'the knowledge-based economy' (David & Foray, 1995; OECD, 1996b). This led to the yearly publication of the so-called *Science, Technology, and Industry Scoreboards*,[3] and the periodic summary of progress at the ministerial level in *Science and Technology Statistical Compendia*.[4] However, Godin (2006b, at p. 24) evaluated that the metaphor of a 'knowledge-based economy' has functioned, in this context, mainly as a label for reorganizing existing indicators—most of the time, assuming national systems of member states explicitly or implicitly as units of analysis. He warned again that "important methodological difficulties await anyone interested in measuring intangibles like knowledge" (p. 24).

## 2. The Triple Helix as a model of the knowledge-based economy

In the Triple Helix model of the knowledge-based economy, the main institutions have first been defined as university, industry, and government (Etzkowitz and Leydesdorff, 1995). However, these institutional carriers of an innovation system can be expected to entertain a dually layered network: one layer of institutional relations in which they constrain each other's behavior, and another layer of functional relations in which they shape each other's expectations. For example, the function of university-industry relations can be performed by different institutional arrangements such as transfer offices, spin-off companies, licensing agreements, etc. The institutional relations provide us with network data, but the functions in a knowledge-based economy are to be analyzed in terms of the transformative dynamics. The knowledge base of an economy can be considered as a specific configuration in the structure of expectations which feeds back as a transformation mechanism on the institutional arrangements.

---

[3] The *Science, Technology, and Technology Scoreboard 2007* is available at
http://www.oecd.org/document/10/0,3343,en_2649_33703_39493962_1_1_1_1,00.html.
[4] OECD Science, Technology and R&D Statistics Online Database is available at
http://www.oecd.org/document/50/0,3343,en_21571361_33915056_39132274_1_1_1_1,00.html; The
*Science and Technology Statistical Compendium 2004* is available at
http://www.oecd.org/document/8/0,2340,en_2649_33703_23654472_1_1_1_1,00.html

How would a knowledge-based economy operate differently from a market-based or political economy? The market mechanism first equilibrates between supply and demand. Secondly, economic exchange relations can be regulated by political institutions. I shall argue that organized knowledge production has more recently added a *third* coordination mechanism to the social system in addition to economic exchange relations and political control (Gibbons *et al.*, 1994; Schumpeter, [1939] 1964; Whitley, 1984).

Three sub-dynamics are reproduced as functions of a knowledge-based economy: (1) wealth generation in the economy, (2) novelty generation by organized science and technology, and (3) governance of the interactions among these two subdynamics by policy-making in the public sphere and management in the private sphere. The economic system, the academic system and the political system can be considered as relatively autonomous subsystems of society which operate with different mechanisms. However, in order to describe their mutual interdependence and interaction with respect to knowledge creation, one first needs to distinguish these mechanisms.

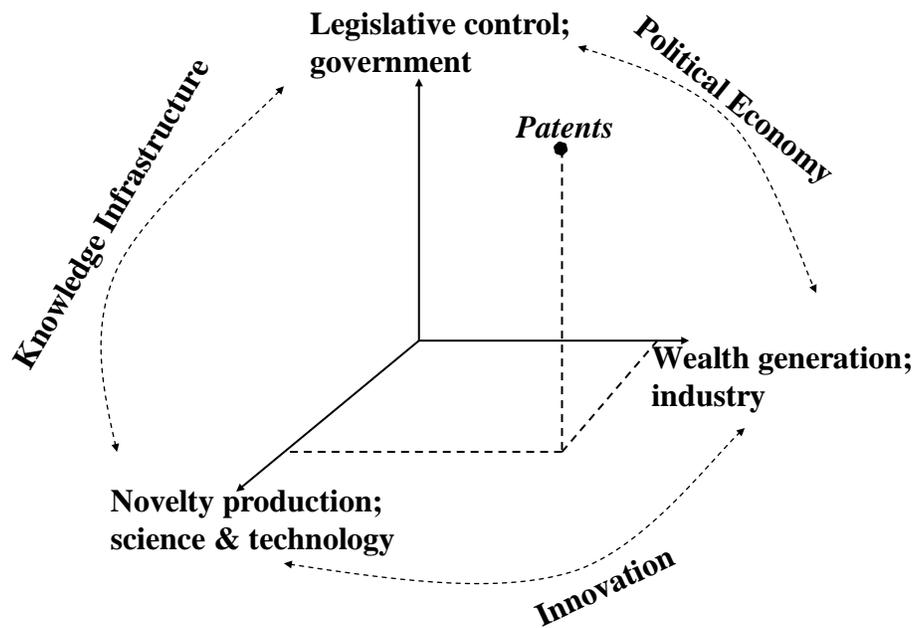

**Figure 1**: Patents as events in the three-dimensional space of Triple Helix interactions.

The three sub-dynamics are not given, but constructed and continuously reconstructed in social relations. They can be considered as three helices operating upon each other selectively. For example, a patent can be considered as an event in which the coordination mechanisms interact (Figure 1).[5] The interactions among these functionally differentiated mechanisms drive a cultural evolution which requires a model more complex than the biological model of evolution (Luhmann, 1984 [1995a]; 1990, at p. 340).

Biological evolution assumes 'natural selection' as a single selection mechanism. When selecting recursively from each other, two selection mechanisms can be expected to develop into a co-evolution—as in a process of 'mutual shaping.' The dynamics among three selection mechanisms, however, can be expected to lead to a higher degree of non-linearity and therefore complexity (Li and Yorke, 1975; May & Leonard, 1975; May, 1976). The resulting complex dynamics evolves in terms of trajectories and regimes that change the system in which they emerge (Dosi, 1982).

In such a complex dynamics, the independent (steering) variables at one moment of time may become dependent at a next moment. In other words, the dynamics can become self-organizing because the incentives for change can come from different sources and more than a single selection environment is operating. Consequently, the economic and political mechanisms no longer control, but function as selective feedback mechanisms that enable and constrain the development of scientific and technological knowledge. *Mutatis mutandis*, the development of scientific and technological knowledge has become a structural condition and a limiting factor on further socio-economic development.

The analytical function of the Triple Helix model is to unravel the complex dynamics of a knowledge-based economy in terms of its composing subdynamics. The formal model is not a grand super-theory: it builds on and remains dependent on appreciations of the phenomena at the level of the composing theories. Developments in the various discourses provide the data. Not incidentally, the Triple Helix model originated from the study of science and technology (Etzkowitz & Leydesdorff, 2000; Mirowski & Sent, 2007; Shinn, 2002; Slaughter & Rhodes, 2004).

The model is not first specified in terms of domains (e.g., national systems) or specific functions (e.g., knowledge production), but allows for interaction effects among domains and specific synergies among functions and institutions. The various subdynamics in the model can be considered from different analytical perspectives. These perspectives may be perpendicular and therefore develop as incommensurable discourses. Neo-classical economics, for example, has focused on the market as an equilibrating mechanism at each moment of time, whereas evolutionary economics focuses on innovations that upset the equilibria over time (Nelson & Winter, 1982). However, the different perspectives can be reconstructed in terms of their contributions to the specification of a Triple Helix model of the knowledge-based economy. Reflection on and abstraction from the specification of substantive mechanisms in favor of formalization enables us to understand the knowledge base of a social system as a complex dynamics of expectations which are specified in variously coded communications.

---

[5] The different mechanisms are integrated in the observable events (in this case, the patents) as co-variations among the latent dimensions. However, the dimensions are different in terms of their remaining variations. The variations in each dimension (that is, the covariation plus the remaining variations) develop over time recursively, that is, with reference to a previous state.

Political discourse, for example, can be expected to operate differently from scientific discourse. Insofar as an overlay system of translations among these analytically different codes of communication can be shaped and reproduced, a next-order level of expectations may be stabilized and induce globalization as an additional feedback at the systems level. This feedback itself, however, remains a fallible expectation. The differentiations and translations among the variously codified expectations drive the development of the knowledge base in a social system.

In other words: three interacting subdynamics can be expected to generate hypercycles on top of the cycles (e.g., business cycles, election cycles, paradigm changes) between the constitutive subdynamics. Whereas structures in the data develop along trajectories, the hypercycle provides us with a next-order regime. This regime may be more or less knowledge-based depending on the configurations among the selection mechanisms on which it builds at the structural level.

For example, as long as economic exchanges and political control are not systematically affected by knowledge production and control, a political economy can be expected to prevail (Richta *et al.*, 1968). In a knowledge-based economy, however, three levels can be distinguished: the level of the data where information is exchanged in (e.g., economic) relations, the level of (e.g., institutional) structures operating selectively—at this level specific meaning can be codified and selectively exchanged—and a third level at which configurations of meaning-exchanges can be knowledge-based to varying extents.

In summary, this model is differentiated both horizontally and vertically. Horizontally different coordination mechanisms operate upon one another. Vertically, the information is structured semantically and the structures can develop along trajectories, while the trajectories are embedded in regimes that emerge from configurations among structures and trajectories. Luhmann (1984 [1995a]; 1997) distinguished the horizontal differentiation as *functional* differentiation from the vertical one as *social* differentiation. Horizontal differentiation is based on differences among the codes of communication in the coordination systems, while vertical differentiation corresponds with the distinction between institutional and functional dynamics in the Triple Helix model. Different subdynamics can be expected to operate in different layers, and to interact in the instantiations.

At the institutional level, functional dynamics are integrated historically in one arrangement or another, and the various subsystems are thus instantiated (Giddens, 1984). These historical instantiations condition and enable the further development of functional subsystems which develop in terms of flows of communication. Analytically, the instantiations enable us to specify the dynamics in terms of relevant parameters. The selection mechanisms, however, remain theoretical specifications with the epistemological status of a hypothesis. Entertaining the Triple Helix hypothesis that more than two selection environments need to be specified for the analysis of manifestations of the knowledge-based economy can enrich both the description and the analysis.

For example, in the context of a knowledge-intensive corporation, technological expectations have to be combined with market opportunities in planning cycles. Decisions in interactions at the organizational level shape the future orientation of the corporation, but have only localizable effects on the global market and the relevant technologies (Luhmann, 2000). These other two helices are instantiated by the decision in the third one. However, the instantiations should not be confused with the dynamics.

In university-industry relations, for example, the institutional arrangements (e.g., transfer offices) can be evaluated in terms of how well they serve the transfer of new knowledge in exchange for university income. The knowledge transfer process may be enhanced or hindered by the institutional contexts. Thus, both the functions and the institutions remain contingent. Furthermore, the two layers can be coupled operationally; they may co-evolve or not. In the case of a co-evolution this may lead to a dead end ('lock-in') or provide a competitive advantage. These remain empirical questions.

In a pluriform society, the processes at different levels and in different dimensions develop concurrently, asynchronously, and in interaction with one another. The horizontal differentiation among the coordination mechanisms is based on the availability of different codes for the communication. Because meanings are no longer given, but constructed and reproduced in discourses, the codes can also be translated into one another. For example, the political system can (attempt to) regulate the market because it first uses a code different from that of the market mechanism. However, the political system can, in addition to legislation, also reconstruct the market by creating market incentives. The codes are historically recombined by local agency, including institutional agency.

Insofar as the three (latent) functions resonate into a configuration, a knowledge-based economy can be generated. The various 'horizons of meaning' (Husserl, [1929] 1973, p. 45, pp. 133f.) in the different dimensions resound in the communicative events which instantiate these systems. The observable instantiations can be expected to change the expectations in a next round. Note that there may be more than one configuration which is knowledge-based. In other words, the knowledge-basedness of a system poses an empirical question. The knowledge-based economy can be considered as a configuration among three leading coordination mechanisms such that a degree of freedom is added to the two previously leading mechanisms in a political economy (markets and political control).

While the market provides analytically only a single selection mechanism, two selection mechanisms can reinforce each other into the cycles of a political economy. A system with three selection mechanisms can additionally be globalized on top of the institutional stabilizations. Globalization, however, remains a construct (with the theoretical status of a hypothesis about a hyper-cycle operating at a next-order level). The reproduction of this next-order system cannot be controlled by the constructing system because a self-organizing dynamics is supposedly generated.

In other words, globalization at the regime level remains a tendency in the historical systems (trajectories) which are globalizing. The two layers of institutional stabilization and retention *versus* functional restructuration are both needed and can be expected to feed back into each other, thus changing the institutional roles, the selection environments, and potentially the evolutionary functions of the various stakeholders in subsequent rounds of 'creative destruction,' transition, and change (Schumpeter, 1943, at pp. 81 ff.).

The observable networks function as infrastructures of the social system by constraining and enabling communication through them and among them.[6] Insofar as the fluxes of communication through these networks are guided by the different codes of communication, the system can self-organize a knowledge base in terms of a configuration among the functional subsystems of communication. Because the coordination mechanisms use different rules for the coding, a variety of meanings can be provided to the events from different perspectives. The more the codes of communication can be spanned orthogonally the more complexity can be processed. This differentiation process is counterbalanced by processes of integration in the historical organization by instantiating agency. Thus, the two layers of the Triple Helix model support both differentiation and integration. Trade-offs between globalization and stabilization are endogenous to a knowledge-based economy. Knowledge-based systems remain in transition.

## 3. Knowledge as a social coordination mechanism

Knowledge enables us to codify the meaning of information. Information can be more or less meaningful given a perspective. However, meaning is provided from a system's perspective and with hindsight. Providing meaning to an uncertainty (that is, Shannon-type information) can be considered as a first codification. Knowledge enables us to discard some meanings and retain others in a (next-order) layer of discursive codifications. In other words, knowledge can be considered *as a meaning which makes a difference*.[7] Knowledge enables us to translate one meaning into another. Knowledge itself can also be codified, and codified knowledge can, for example, be commercialized.

---

[6] Whereas the networks of institutional relations can be analyzed by using multi-variate statistics, the functions develop over time. The flows of communication change the networks of communication and a calculus is needed for the analysis of such complex dynamics. Bar-Hillel (1955) noted that Shannon's information theory provides us with such a calculus. This calculus can be elaborated into a non-linear dynamics of probabilistic entropy (Abramson, 1963; Theil, 1972; Brooks & Wiley, 1986; Leydesdorff, 1995 and 2008; Ulanowicz, 1997; Jakulin & Bratko, 2004; Yeung, 2008), which enables us to decompose the complex dynamics in terms of different subdynamics and to specify the various contributions to the prevailing uncertainty in terms of bits of information. Eventually, this information-theoretical approach enables us to measure the knowledge-base of an economy in terms of the Triple Helix model (e.g., Leydesdorff & Fritsch, 2006).

[7] Bateson (1972, at p. 453) defined information as "a difference which makes a difference." From the perspective of this paper, one can consider this as the definition of "meaningful information," whereas Shannon-type information can be considered as a series of differences (Hayles, 1990).

Thus, a knowledge-based system operates in recursive loops of codification that one would expect to be increasingly specific in terms of the information to be retained.

Knowledge informs expectations in the present on the basis of previous operations of the system. Informed expectations orient the discourse towards future events and possible reconstructions. A knowledge-based economy is driven more by reflexively codified expectations than by its historical conditions (Lundvall and Borras, 1997). The knowledge base of a social system can be further developed over time by ongoing processes of theoretically informed deconstructions and reconstructions (Cowan *et al.*, 2000; Foray, 2004).

In other words, science-based representations of possible futures (e.g., 'competitive advantages') feed back on historically manifest processes (Nonaka and Takeuchi, 1995; Biggiero, 2001). The manifestations, however, can be considered as the products of (interactions with) other subdynamics. The other subdynamics (markets, organizations) reflexively counteract upon and thus buffer against the transformative power of the knowledge base. Graham & Dickinson (2007), for example, noted that political discourse may be particularly resistant against the idea that it itself is only one among other subdynamics in a knowledge-based economy.

The reflexive orientation towards the future inverts the time axis locally (Figure 2). A movement against the axis of time is expected to reduce uncertainty.[8] However, a reflexive inversion of the arrow of time may also change the historical dynamics in a historically stabilized system (cf. Giddens, 1990). While stabilization and destabilization can be considered as historical processes (along the axis of time), reflexivity adds to the historical process an evolutionary dynamics that is based on selections in the present from the perspective of hindsight. By inverting the axis of time, stabilizing dynamics can under certain conditions also be globalized (Coveney and Highfield, 1990; Mackenzie, 2001; Urry, 2000; 2003). Reflexivity, however, enables us to develop both perspectives as two sides of the same coin: the historical instantiations along trajectories and the evolutionary dynamics of expectations at the regime level.

---

[8] It can be shown that the Second Law holds equally for probabilistic and thermodynamic entropy production (Theil, 1972, at pp. 59 ff.).

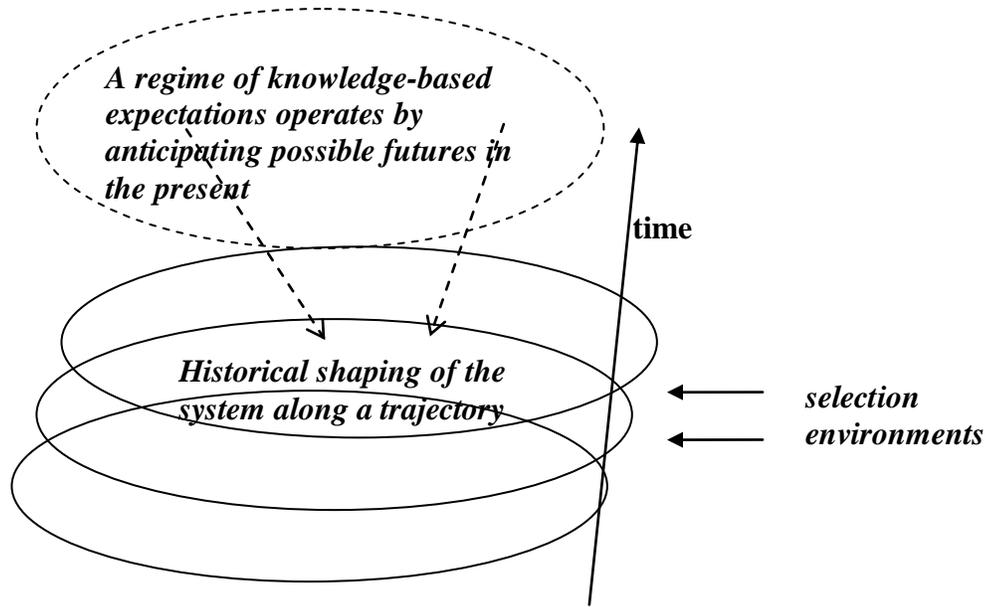

**Figure 2**: A technological trajectory follows the axis of time, while a knowledge-based regime operates within a system in terms of expectations, that is, against the axis of time.

Let us first follow the construction of a knowledge-based economy from the historical perspective. Codified knowledge has always been a relevant source of variance at the level of individual actions. Before the emergence of scientific and technological knowledge as another mechanism of social coordination, the economic exchange of knowledge was developed as distinct from the exchange of commodities within the context of the market economy.

For example, the patent system can be considered as a typical product of industrial competition. Patent legislation became crucial for regulating intellectual property in the late nineteenth century when knowledge markets emerged in chemistry and later in electrical engineering (Noble, 1977; Van den Belt and Rip, 1987). Patents package scientific knowledge so that new knowledge can function at the interface of science with the economy and be incorporated into knowledge-based innovations (Granstrand, 1999; Jaffe and Trajtenberg, 2002). Patents thus provide a format for codifying knowledge for purposes other than the internal requirements of quality control in scientific communication.

The organized production and control of knowledge for the purpose of industrial innovation increasingly emerged as a subdynamic of the socio-economic system in advanced capitalist societies since approximately 1870 (Braverman, 1974; Noble, 1977). Schumpeter ([1939], 1964) is well-known for his argument that the dynamics of innovation upset the market mechanism (Nelson and Winter, 1982). While market mechanisms seek equilibrium at each moment of time, novelty production generates an orthogonal subdynamic along the time axis. In economics, this has been modeled as the difference between factor substitution (the change of input factors along the production function) and technological development (a shift of the production function towards the origin; Sahal, 1981a and b). Technological innovations enable enterprises to reduce factor costs in both labor and capital (Salter, 1960; cf. Rosenberg, 1976).

Improving a system innovatively presumes that one is able to handle the system purposefully. When this reflection is further refined by organizing knowledge, the innovative dynamics can be reinforced. This reinforcement will occur at some *places* more than at others. Thus, a third dimension pertinent to our subject can be specified: the geographical—and potentially national—distribution of whatever is invented, produced, traded, or retained. Nation-states, for example, can be expected to differ in terms of the relationship between their respective economy and its knowledge base (Lundvall, 1992; Nelson, 1993).

Geographically positioned units of analysis (e.g., firms, institutions), economic exchange relations, and novelty production cannot be reduced to one another. However, these independent dimensions can be expected to interact to varying extents (Storper, 1997). Given these specifications, one can create a model of the three dimensions and their interaction terms as follows:

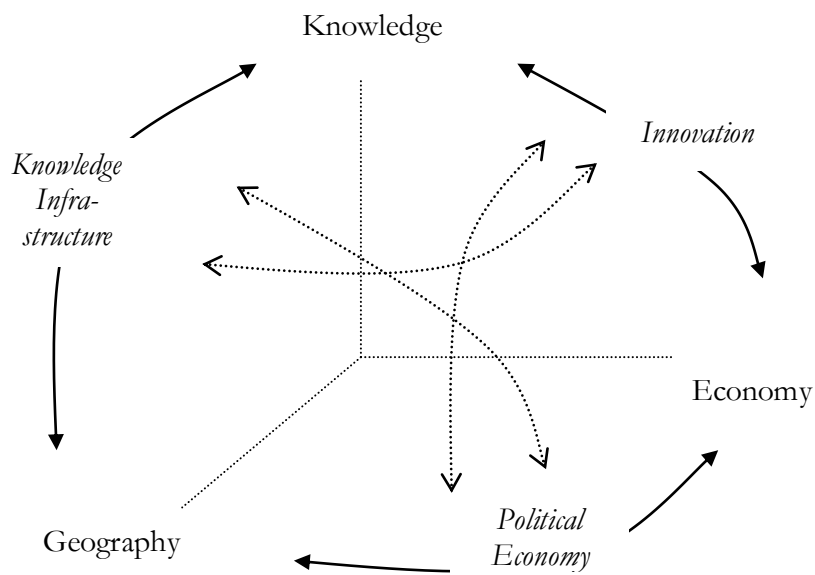

**Figure 3:** Three dimensions of the social system with their three interaction terms.

This distinction of three dimensions will enable me in a later section to specify different micro-operations of the social system because agents (1) are geographically positioned and therefore locally embedded, (2) can maintain economic exchange relations across borders, and (3) learn from the resulting dynamics with reference to their positions and relations. The first micro-operation is considered in neo-classical economics as the micro-foundation of individual agency: agents are considered as *endowed* with natural preferences. The second micro-operation of *interaction* has been proposed by Lundvall (1988) as the alternative foundation of his program of national systems of innovation in evolutionary economics. I return in more detail to this proposal below. The reflexive layer, however, adds a *third* micro-operation: reflection can be both individual and relational. We shall see below that this latter distinction also makes a difference: *reflexive communications* can be expected to transform networks of relations in an evolutionary mode (Beck *et al.*, 1994; Bhaskar, 1979; 1998, at p. 207; Maturana, 1978; Leydesdorff, 2009).

Figure 3 elaborates the conceptualization by situating each of the interaction terms between two of the three dimensions. A political economy, for example, organizes markets within the context of a nation-state. The third coordination mechanism of knowledge and learning is more recently added and may change a political economy into a knowledge-based one (at some places more than other). In a pluriform and differentiated society, however, the various interaction terms are no longer synchronized *ex ante*, and thus they may begin to interact among themselves.

The developments are not coordinated and may thus develop in some dimensions at some places more than at other. For example, military anticipations which primarily developed in national contexts shape also non-market selection environments which enable firms to move forward in constructing competitive advantages in a relatively shielded environment (Rosenberg, 1976 and 1990). Thus, the Triple Helix dynamics is not developed at a meta-level, but endogenous once this model of sub-dynamics operating upon one another is available within a society. The social organization of knowledge-based expectations feeds back as a regime by transforming the trajectories from which it emerged.

## 4. Neo-evolutionary dynamics in a Triple Helix of coordination mechanisms

During the formation of political economies in national systems during the nineteenth century, knowledge production was at first considered as an exogenous given (List, 1841; Marx, 1867).[9] Under the condition of constitutional stability in the various nation-states after approximately 1870,[10] *national systems of innovation* could gradually be developed among the axes of economic exchange and organized knowledge production and control (Noble, 1977; Rosenberg, 1976, 1982a). Globalization, however, restructures the relations among nations. The variation among nations provides another dimension to the global system. Interactions among three subdynamics, however, can be expected to generate a complex dynamics of transition.

---

[9] Marx (1857) closely observed the *technological* condition of industrial capitalism, noting, for example, that: "Nature does not build machines, locomotives, railways, electric telegraphs, self-acting mules, etc. These are the products of human industry; natural resources which are transformed into organs of the human control over nature or one's practices in nature. (…) The development of fixed assets shows the extent to which knowledge available at the level of society is transformed into immediate productive force, and therefore, the extent to which the conditions of social life have themselves been brought under the control of the general intellect and have been transformed accordingly. Crucial is the degree to which the socially productive forces are produced not only as knowledge, but as immediate organs of social practice, that is, of the real process of living" (Marx, 1857, at p. 594; my translation). Although reflexively aware of the potential dynamics in organized knowledge production and control, Marx thus remained focused on the historical state of the development of science and technology, and on the integration of this condition into the political economy.

[10] In 1870, Germany and Italy were unified; France had gone through a revolution leading to the establishment of a modern (third) republic. The Meji Restoration of 1869 had made Japan a player in the industrial competition, and the U.S.A. had emerged from the Civil War in 1865. After approximately 1870, the economic system had been reshaped into a system of nations with their respective political economies.

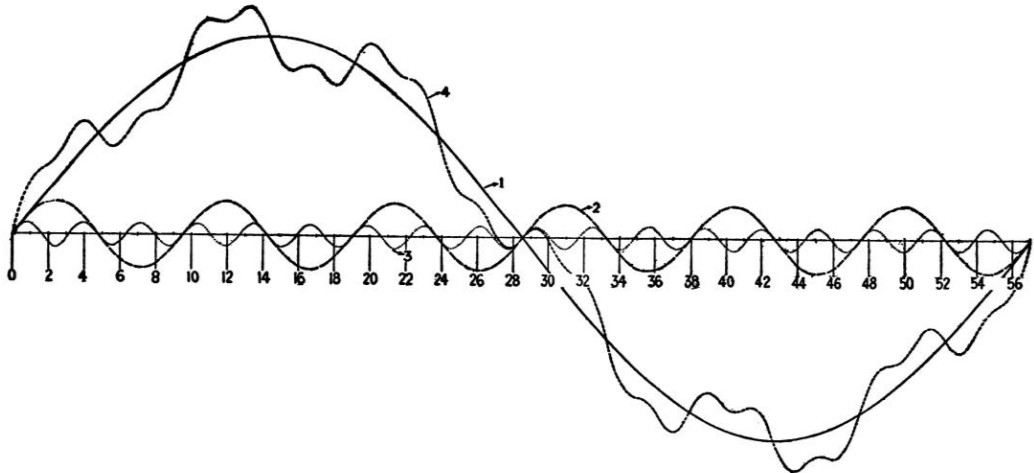

CHART I.—Curve 1, long cycle; curve 2, intermediate cycle; curve 3, short cycle; sum of 1–3.

**Figure 4**: Interaction among three regular cycles can lead to a chaotic pattern. Source: Schumpeter ([1939] 1964, at p. 175).

It is less known that Schumpeter ([1939] 1964, at p. 175) himself signaled already that the superposition of three cycles can be expected to generate chaotic patterns (Figure 4). In general, two interacting subdynamics can be expected to co-evolve along trajectories as long as the third dynamic can be considered as relatively constant. Over time, two subdynamics can also lock-in into each other in a process of mutual shaping in a co-evolution (Arthur, 1994; McLuhan, 1964). However, stabilities or regular patterns developed between two subdynamics can be de-stabilized by a third. Historically, a hitherto stable context may begin to change under the pressure of an emerging subdynamic. For example, the erosion of relative stability in nation-states because of more recent globalization processes has changed the conditions of national innovation systems.

While a political economy can be explained in terms of two subdynamics (for example, as a 'dialectics' between production forces and production relations), a complex dynamics can be expected when three subdynamics are left free to operate upon one another. However, a configuration with three possible degrees of freedom—markets, governance, and knowledge production—can be modeled in terms of a Triple Helix of university-industry-government relations (Etzkowitz and Leydesdorff, 2000; cf. Lewontin, 2000). Governance can be considered as the variable that instantiates and organizes systems in the geographical dimension of the model, while industry is the main carrier of economic production and exchange. Thirdly, universities play a leading role in the organization of the knowledge-production function (Godin and Gingras, 2000).

In this (neo-)evolutionary model of interacting subdynamics, the institutional dimensions can no longer be expected to correspond one-to-one with the functions in the networks carried by and between the agencies. The analytical dimensions span a space in which the institutions operate. Each university and industry, for example, also has a geographical location and is therefore subject to regulation and legislation. In a knowledge-based system, however, functions no longer develop exclusively at the local level, that is, contained within institutional settings. Instead, the interactions generate evolutionary dynamics of change in relations at the network level. In other words, the functions provide a layer of development that is analytically different from, but historically coupled to the institutional arrangements.

Two of the three functions (economy and science) can be considered as relatively open since 'universal' (Parsons, 1951; Luhmann, 1984 [1995a], 1990). The function of control bends the space of possible interactions reflexively back to the *positions* of the operating units (e.g., the firms and the nations) in the marketplace and at the research front, respectively. In this dimension, the question of what can be retained locally during the reproduction of the innovation processes becomes crucial. The system of reference for this question about how to organize and interface the fluxes of communication remains the political economy.

Government policies and management strategies weave a reflexive layer with the private/public distinction as its specific degree of freedom. The advantages of entertaining a knowledge base can only be incorporated if the knowledge produced by the interacting fluxes can also organizationally and reflexively be retained by this network. In other words, the development of a knowledge base is dependent on the condition that knowledge production be socially organized into a knowledge infrastructure; for example, in R&D laboratories. The historical development of a regime of intellectual property rights reflects the function of government policies in securing the knowledge-based dynamics (Kingston, 2003).

As noted, institutions and functions can be expected to co-evolve in some configurations more than in others. However, these co-evolutions can continuously be disturbed by a third subdynamic. The knowledge-based economy cannot be developed without the consequent destabilizations and reconstructions, or, in Schumpeters (1943, at pp. 81 ff.) terminology, 'creative destruction.' The destabilizing dynamics of innovation can be reinforced when they are knowledge-based because the dynamics of reconstruction are reinforced (Barras, 1990; Freeman & Perez, 1988). Knowledge can make alternatives available, and the more codified this knowledge is, the more globally this communication system can function as another selection context.

The expectations which are organized in a knowledge base can further be codified like in scientific knowledge. The expectations can also be codified through the local use of knowledge and knowledge can be retained in textual practices. When increasingly organized by R&D management and S&T policies, the emerging structures in the expectations provide another selection environment. This global selection environment of scientific and technical knowledge remains pending as additional selection pressure upon the locally stabilized configurations. Thus, codification both stabilizes and globalizes discursive knowledge, and the third subdynamic of knowledge-based innovations can become increasingly a global driver of change by facilitating these changes of perspectives.

For example, Dosi (1982) distinguished between the stabilization of innovations and routines along technological *trajectories* and the knowledge base as a next-order *regime* that remains emergent as a paradigm. As innovations are further developed along trajectories, a knowledge base becomes reflexively available as an evolutionary mechanism for restructuring the historical trajectories on which it builds. This next-order perspective of the regime rests as an additional selection environment on the historical trajectories. In terms of the previous figure, this second-order system can be added as in Figure 5.

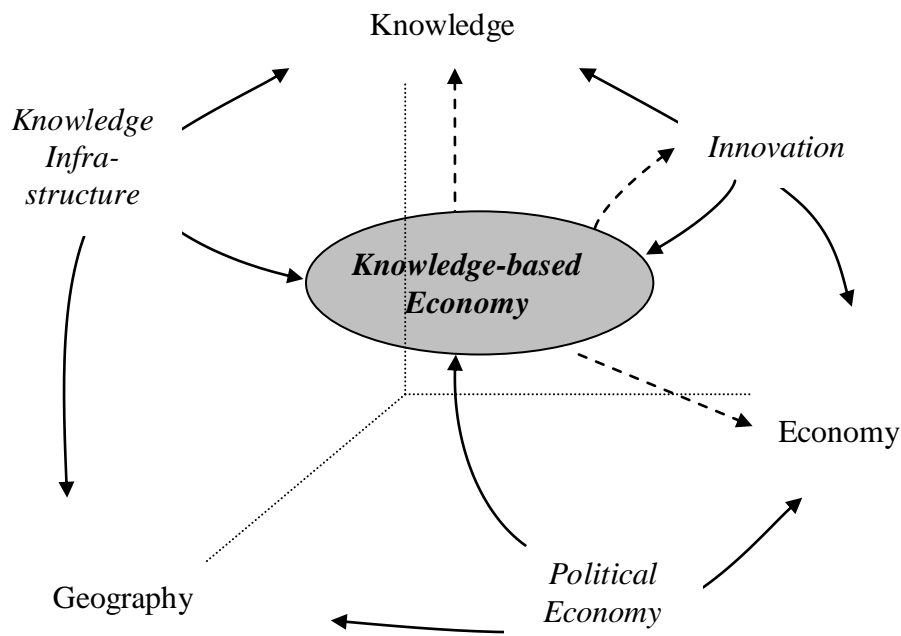

**Figure 5:** The first-order interactions generate a knowledge-based economy as a next-order system.

While the construction of a knowledge base remains a bottom-up operation, a control mechanism that tends to operate top-down, is increasingly constructed at the systems level. DiMaggio & Powell (1983, at p. 150) called this emerging control 'coercive isomorphism', and Giddens (1979, at pp. 77 ff.) specified the mechanism as a 'duality of structure' which, according to him, would operate in a 'virtual' dimension (*ibid.*, p. 64). However, these specifications remained at the level of behaviour and action, and did not specify the hypothesized dynamics as the evolutionary self-organization of expectations at a next-order level. Giddens (1984, at p. *xxxvii*) deliberately abstained from the specification of next-order control mechanisms because it would in his opinion lead to reintroducing and potentially reifying a systems perspective à la Parsons (cf. Merton, 1973).

From a systems perspective, Giddens's warning entails a *non sequitur*. Reification does not follow from an analytical distinction. The analytical distinction remains part of a discourse. However, the exclusive focus on observable variation in the action-oriented approaches and from the neo-institutional perspective prevents us from further developing an evolutionary and sociological theory of communications at more abstract levels. More abstract theorizing—for example, the specification of expectations about how expectations operate—is needed for understanding how the use of language and discursive knowledge can *change* structure/action contingency relations.

In my opinion, different selection mechanisms and levels can be distinguished. Note that the dynamics of selection environments are structural and therefore deterministic, although communication systems develop in terms of distributions and therefore uncertainties can be expected in the (distributions of) instantiations of the systems. The proposed change of perspective is analytical: the instantiations (observed values) can inform the expectations, but only insofar as the latter are specified. From this neo-evolutionary perspective, action can be considered as variation, and the focus remains on the specification of the hypothesized selection mechanisms and their interactions.

The codes of communication in these coordination mechanisms provide different meanings to the distributions. A further development of 'the knowledge base of an economy' as a construct can be expected under the condition that the various coordination mechanisms develop specific resonances in interactions among recurring variations, that is, in a self-organizing mode. However, this *self-organization* among the coordination mechanisms of society exhibits an evolutionary dynamics among the flows of communications analytically to be distinguished from the historical *organization* of relations among the carrying institutions (Luhmann, 1975; 1997). While the neo-institutional perspective focused on networks of relations, a knowledge-based economy can only be specified from this neo-evolutionary perspective. The dynamics are based on changes in both the relations and the (potentially meaningful) organization of relations.

## 5. The operation of the knowledge base

Interacting expectations can provide a basis for changes in the behavior of the carrying agents. These behavioral changes differ from the institutional imperatives and market incentives that have driven the system previously (e.g., in terms of profit-maximizing behavior). While institutions and markets develop historically, that is, with the arrow of time, the knowledge-based structures of expectations drive the system in an anticipatory mode. Future-oriented planning cycles can be expected to become (at some places) more important than current prices on the market. Thus, informed anticipations increasingly change the dynamics of the system from an agent-based perspective towards a more abstract knowledge-based one.

The knowledge-based subdynamic operates by reconstructing the past in the present on the basis of representations that contain informed expectations (e.g., curves and functions on computer screens). As the intensity and speed of communication among the carrying agencies increases (e.g., because of the further developments of ICT; cf. Kaufer & Carley, 1993; Luhmann, 1997, pp. 303 ff.), the codification of knowledge becomes increasingly a functional means to reduce complexity in inter-human communication. The emerging order of expectations remains accessible to reflexive agents with a capacity to learn. The expectations can be refined as they become more theoretically informed. The communicative competencies of the agents involved can be considered as rate-limiting factors.

Both participants and analysts are able to improve their understanding of the restructuring of the expectations at interfaces within the systems under study, and one is able to switch roles albeit to variable extents (Giddens, 1976). As these communicative competencies are further developed among the carriers of the communications, the codifications in the expectations can be further developed. For example, in a knowledge-based economy the already abstract price-mechanism of a market-based economy can increasingly be reconstructed in terms of price/performance ratios based on expectations about the life-cycles of technologies (Galbraith, 1967; Heertje, 1973). No educated consumer nowadays buys a computer, for example, using only price comparisons. Thus, more abstract and knowledge-intensive criteria are increasingly guiding economic decision-making. These more abstract criteria have been central to government (e.g., procurement) policies in advanced industrial nations since the early 1980s in response to the second oil crisis (Freeman, 1982; Rothwell & Zegveld, 1981).

The dynamics of a complex system of knowledge-based innovations are non-linear (Allen, 1994; Krugman, 1996). This non-linearity is a consequence of interaction terms among the subsystems and the recursive processes operating within each of them simultaneously. In the long run, the non-linear (interaction) terms can be expected to outweigh the linear (action) terms because of the higher exponents in the equations. For example, the *interaction* effects between 'demand pull' and 'technology push' can over time be expected to become more important for the systemic development of innovations than the sum of the linear action terms (Kline and Rosenberg, 1986; Mowery and Rosenberg, 1979, 1989). The non-linear interaction terms can be expected to lead to 'unintended consequences.'

For example, when a sector undergoes technological innovation, a 'lock-in' into a market segment may first shape a specific trajectory of innovations (David, 1985; Arthur, 1994; cf. Liebowitz & Margolis, 1999). Learning curves can be steep following such a breakthrough and the stabilization of a trajectory in the marketplace (Arrow, 1962; Rosenberg, 1982b). Analogously, when a science-based technology locks into a national state (e.g., in the energy or health sector), a monopoly can be immunized against the third helix of market forces for considerable periods of time. Over longer periods of time, however, these lock-ins can be expected to erode because of the ongoing processes of 'creative destruction' (Schumpeter, 1943). Such creative destruction is based on recombinations of market forces with new insights. For example, the monopolies of national telephone companies in Western Europe disappeared under the pressure of new (neo-liberal) industrial policies and the emergence of the cell phone. Corporations may have to disinvest in their own competencies in order to make room for new technologies.

For example, when the DC-3 was introduced in the 1930s, it took Boeing two years to construct a competing airplane (the 307 Stratoliner) on the basis of adopting the 'textbook' of its main competitor. When in a next revolution Boeing took the lead with the introduction of wide-body airplanes (the 707-series) in 1957, it took the company almost 25 years to establish its model as the lead paradigm because in the meantime competition had become global and the old paradigm of propeller airplanes was well-established. Only when Airbus adopted the new paradigm in 1981 did the textbook of the jet-engine powered airplane become fully established (Frenken, 2005; Frenken & Leydesdorff, 2000). Old trajectories can survive in niche markets and this may postpone crises. Interaction effects among negative feedbacks, however, may lead to global crises that require the restructuring of the carrying layer of institutions (Freeman and Perez, 1988).

Historically, interactions among the relevant (sub)dynamics were first enhanced by geographical proximity (for example, within a national context or the context of a single industry), but as the economic and technological dimensions of the systems globalized, dynamic scale effects became more important than static ones for the retention of wealth. Such dynamic scale effects through innovation were first realized by multinational corporations (Galbraith, 1967; Granstrand *et al.*, 1997; Brusoni *et al.*, 2000). They became a concern of governments in advanced, industrialized countries after the (global) oil crises of the 1970s (OECD, 1980). Improving the knowledge base in the economies of these nations became a priority as science-based innovations were increasingly recognized as providing the main advantages to these economies (Freeman, 1982; Irvine and Martin, 1984; Porter, 1990; Rothwell and Zegveld, 1981).

In other words, the relatively stabilized arrangements of a political economy endogenously generate the meta-stability of a knowledge-based system when the geographical units begin to interact and exchange more intensively in the economic and technological dimensions. Under the pressure of globalization, the institutional make-up of the national systems must be restructured: the national and international perspectives induce an 'oscillation' in a system between its stabilized and globalized states. The oscillating system uses its institutional resources (among which its innovative capacities) for the continuation of an endless transition.

## 6. The restructuring of knowledge production in a KBE

The knowledge base emerges by recursively codifying the expected information content of the underlying arrangements (Maturana and Varela, 1980; Fujigaki, 1998; Luhmann, 1984 [1995a]). A cultural 'reality' of expectations is constructed piecemeal on top of a reality which seems 'naturally given' (Berger and Luckman, 1966). For example, what is natural or sustainable in a Dutch polder landscape can be defined differently with reference to different centuries in the past. 'Nature' is no longer a given, but can be specified as one of the previous states of a technology. However, the technological expectations change with time. In a knowledge-based economy this reconstruction of expectations has itself become the systemic consequence of industrial production.

Wide-body airplanes, for example, make mass-tourism possible to destinations that were previously out of reach for large parts of the population. The regime level of expectations ('a winter break on a tropical island') emerges from the previous trajectories (e.g., of wide-body airplanes), but as a socially unintended consequence and a new market. A previously stabilized system globalizes with reference to its next-order or regime level as an order of expectations.

Innovations can be considered as the historical carriers of this transformative regime because they reconstruct, reorganize, and thus recontextualize the relevant interfaces among the relevant selection environments. Innovations instantiate the globalizing dynamics in the present and potentially restructure existing interfaces in a competitive mode. In an innovative environment, the existing arrangements have to be reassessed continuously. For example, if one introduces high-speed trains, the standards and materials for constructing railways and rails may have to be reconsidered.

Once in place, a knowledge-based system thus feeds back on the terms of its construction by offering comparative improvements and advantages to the solutions found hitherto, that is, on the basis of previous crafts and skills. Knowledge-intensity drives differentiation at the global level by providing us with alternatives. However, the emerging system continues to operate locally in terms of institutions and solutions that organize and produce observable integration across interfaces. The production facilities provide the historical basis—that is, the knowledge infrastructure—for further developing the knowledge-based operations within it.

Because of this historical shaping under interacting selection pressures of highly codified systems of communication, the expectations remain heavily structured and invested with interests. Finding solutions to puzzles requires investments of time and money. Some authors (Gibbons *et al.*, 1994; Nowotny *et al.*, 2001) have claimed that the contemporary system exhibits de-differentiation among policy-making, economic transactions, and scientific insights due to the mutual 'contextualization' of these processes. These authors posit that a new mode of operation ('Mode-2') has emerged at the level of the social system because of the dynamics of incorporating scientific knowledge.

Indeed, the perpetual restructuring of the system which is guided by the knowledge base (of interacting structures of expectations) can be expected to induce new institutional arrangements. Such rearrangements may include the (perhaps temporary) reversal of traditional roles between industry and the university, e.g., in interdisciplinary research centers (Etzkowitz *et al.*, 2000). Among codified expectations, however, exchanges can be expected to remain highly structured and continue to reproduce differentiation for evolutionary reasons: a differentiated system of communications can process more complexity than an integrated one (Shinn, 2002; Shinn & Lamy, 2006). The integration in the instantiations ("de-differentiation") can be made fully compatible with a model that assumes functional differentiation among the codes of communications that flow through the instantiations.

Complex systems need both the local *integration* of the various subdynamics into organizational formats and global *differentiation* among the codes of communication in order to enhance further developments. This tension allows for meta-stabilization as a transitory state that can sustain both innovation and retention. In such systems, functions develop both in interactions with one another and along their own axes, and thirdly in interaction with the exchanges among the institutions. An integrated perspective like 'Mode-2' provides a specific combination of perspectives among other possible ones.

At the interfaces between the economics of the market and the heuristics in R&D processes, different translation mechanisms can be further developed that structure and codify these interactions over time. I cited the example of developing the price mechanism into the (bivariate) price/performance criterion, but in innovative environments one can expect criteria to become multivariate. For example, knowledge-based corporations organize sophisticated interfaces between strategic (long-term) and operational (medium-term) planning cycles in order to appreciate and to update the different perspectives (Galbraith and Nathanson, 1978).

Since communications in a knowledge-based system are no longer controlled by a single coordination mechanism, integration and differentiation can be expected to operate concurrently at the various interfaces, and without *a priori* synchronization. In terms of the dynamics of the system, differentiation and integration can thus be considered as two sides of the same coin: integration may take different forms, and differentiations can be relatively integrated (as subsystems). From an evolutionary perspective, the question becomes: where in the network can the relevant puzzles be solved and hence competitive edges be maintained? Thus, one can expect both geographically confined innovation systems and technological systems of innovation (Carlsson, 2002, 2006; Carlsson and Stankiewicz, 1991; Edqvist, 1997). The horizontal and vertical overlapping of systems and subsystems of innovation can be considered a hallmark of the knowledge-based economy.

In other words, the definition of a system of innovations becomes itself increasingly knowledge-based—that is, a research question!—in a knowledge-based economy, since the subsystems are differently codified, yet interacting (at different speeds) in the reproduction of the system. Governance of a knowledge-based economy can only be based on a set of informed assumptions about the relevant systems (Weiss, 1979). These hypotheses are predictably in need of more informed revisions because one expects new formats to be invented at the hitherto stabilized interfaces.

## 7. The KBE and the systems-of-innovation approach

When Lundvall (1988) proposed that the nation be considered as the first candidate for the *integration* of innovation systems, he formulated this claim carefully in terms of a heuristics:

> The interdependency between production and innovation goes both ways. […] This
> interdependency between production and innovation makes it legitimate to take the
> national system of production as a starting point when defining a system of innovation.
> (Lundvall, 1988, at p. 362)

The idea of integrating innovation into production at the *national* level has the analytical
advantage of providing the analyst with an institutionally demarcated system of reference.
If the market is continuously upset by innovation, can the nation then perhaps be
considered as another, albeit institutionally organized (quasi-)equilibrium (Aoki, 2001)?

The specification of the nation as a well-defined system of reference enabled
evolutionary economists to study, for example, the so-called 'differential productivity
growth puzzle' which is generated by the different speeds of development among the
various industrial sectors (Nelson and Winter, 1975). The problem of the relative rates of
innovation cannot be defined properly without the specification of a system of reference
that integrates different sectors of an economy (Nelson, 1982, 1994). The solutions to this
'puzzle' can accordingly be expected to differ among nation-states (Lundvall, 1992;
Nelson, 1993).

The emergence of transnational levels of government like the European Union, together
with an increased awareness of regional differences within and across nations, have
changed the functions of national governments (Braczyk *et al.*, 1998). The historical
progression varies among countries; integration at the national level still plays a major
role in systems of innovation (Skolnikoff, 1993). However, 'government' has evolved
from a hierarchically fixed point of reference into the variable 'governance' that spans a
variety of sub- and supranational levels. Larédo (2003) argued that this polycentric
environment of stimulation has become a condition for effective innovation policies in
the European Union.

Innovations are generated and incubated by locally producing units such as scientific
laboratories, artisan workshops, and communities of instrument makers, but in interaction
with market forces. While the market can be considered in a first approximation as a
global and relatively open network seeking equilibrium, innovation requires closure of
the network in terms of the relevant stakeholders (Callon, 1998). This provides
innovation with both a market dimension and a technological dimension. The two
dimensions are traded off at interfaces: what can be produced in terms of technical
characteristics encounters what can be diffused into relevant markets in terms of service
characteristics (Frenken, 2005; Lancaster, 1979; Saviotti, 1996). Thus, a competitive
edge can be shaped locally. Such a locally shielded network density can also be
considered as a *niche* (Kemp *et al.*, 1998; Schot and Geels, 2007). Systems of innovation
can be considered as complex systems because they are based on maintaining interfaces
in a variety of dimensions.

Problems at interfaces may lead to costs, but they can be solved more easily within niches
than in their surroundings. Unlike organizations, niches have no fixed delineations. They
can be considered as densities of interfaces in an environment that is otherwise more loosely connected. Within a niche, competitive advantages are achieved by reducing transaction costs (Biggiero, 1998; Williamson, 1985). Niches can thus be shaped, for example, within the context of a multinational and diversified corporation or within a regional economy. In another context, Porter (1990) proposed analyzing national economies in terms of *clusters* of innovations. Clusters may span vertical and horizontal integrations along business columns or across different types of markets. They can be expected to act as systems of innovation that proceed more rapidly than their relevant environments and thus are able to maintain a competitive edge.

National systems of innovation can be expected to vary in terms of their strengths and weaknesses in different dimensions. In the case of Japan (Freeman, 1988), or in comparisons among Latin American countries (Cimoli, 2000), such a delineation may provide heuristics more than in the case of nations participating in the common frameworks of the European Union. Sometimes, the geographical delineation of systems of innovation in niches is straightforward, as in the case of the Italian industrial districts. These comprise often only a few valleys (Beccatini *et al.,* 2003; Biggiero, 1998). The evaluation of a 'system of innovation' can also vary according to the different perspectives of policy making. While the OECD, for example, has focused on comparing national statistics, the EU has had a tendency to focus on changes in the interactions among the member states, for example, in trans-border regions.[11]

For political reasons one may wish to define a system of innovation as national or regional (Cooke, 2002). However, an innovation system evolves, and its shape is therefore not fixed (Bathelt, 2003). While one may entertain the *hypothesis* of an innovation system, the operationalization and the measurement remain crucial for the validation (Cooke and Leydesdorff, 2006). For example, Riba-Vilanova and Leydesdorff (2001) were *not* able to identify a Catalonian system of innovations in terms of knowledge-intensive indicators such as patents and publications despite references to this regional system of innovation prevalent in the literature on the basis of employment statistics (Braczyk *et al.*, 1998).

Belgium provides an obvious example of regional differentiation. The country has been regionalized to such an extent that one can no longer expect the innovation dynamics of Flanders to be highly integrated with the francophone parts of the country. In a study of Hungary's national innovation system, Lengyel & Leydesdorff (2008) found that three different regimes were generated during the transition period with very different dynamics: (1) Budapest and its agglomeration emerged as a knowledge-based innovation system on every indicator; (2) in the north-western part of the country, foreign-owned companies and FDI induced a shift in knowledge-organization; while (3) the system in the eastern and southern part of the country has remained organized in accordance with government expenditures. In the Hungarian case, the national level no longer adds to the synergy among these regional innovation systems. When Hungary entered its transition period after the demise of the Soviet Union it was probably too late to shape a national system of innovations given the concurrent aspiration of Hungary to access to the European Union.

---

[11] The Maastricht Treaty (1991) assigned an advisory role to the European Committee of Regions with regard to economic and social cohesion, trans-European infrastructure networks, health, education, and culture (Council of the European Communities, 1992). This role was further strengthened by the Treaty of Amsterdam in 1997, which envisaged direct consultations between this Committee of Regions and the European Parliament and extended its advisory role to employment policy, social policy, the environment, vocational training, and transport.

One would expect a system of innovations in the Cambridge (UK) region to be science-based (Etzkowitz *et al.*, 2000), while the system of innovations in the Basque country is industrially based and reliant on technology centers that focus on applied research rather than on universities for their knowledge base (Moso and Olazaran, 2002). In general, the question of which dimensions are relevant to the circumstances of a given innovation system requires empirical specification and research (Carlsson, 2006). However, in order to draw conclusions from such research efforts, a theoretical framework is needed. This framework should enable us to compare across innovation systems and in terms of relevant dimensions, but without an *a priori* identification of specific innovation systems. The systems under study provide the evidence, while the analytical frameworks have to carry the explanation of the differences.

It has been argued above that the emergence of a knowledge base requires the specification of at least three systems of reference. Innovations take place at interfaces, and the study of innovation therefore requires the specification of at least two systems of reference (e.g., knowledge production and economic exchanges). In my opinion, the Triple Helix can be elaborated into a neo-evolutionary model that integrates the 'Mode-2' thesis of the new production of scientific knowledge, the study of systems of innovation in evolutionary economics, *and* the neo-classical perspective on the dynamics of the market. In order to specify this model, the three relevant micro-operations have first to be distinguished analytically and in relation to the most relevant theories of innovation and technological development.

## 8. The KBE and neo-evolutionary theories of innovation

### 8.1 The construction of the evolving unit

Nelson & Winter's (1982) trajectory approach has been central to evolutionary economics. In their seminal study entitled 'In search of useful theory of innovation,' Nelson and Winter (1977) formulated their research program as follows:

> Our objective is to develop a class of models based on the following premises. First, in contrast with the production function oriented studies discussed earlier, we posit that almost any nontrivial change in product or process, if there has been no prior experience, is an innovation. That is, we abandon the sharp distinction between moving along a production function and shift to a new one that characterizes the studies surveyed earlier. Second, we treat any innovation as involving considerable uncertainty both before it is ready for introduction to the economy, and even after it is introduced, and thus we view the innovation process as involving a continuing disequilibrium. […] We are attempting to build conformable sub-theories of the processes that lead to a new technology ready for trial use, and of what we call the

> selection environment that takes the flow of innovations as given. (Of course, there
> are important feedbacks.) (Nelson and Winter, 1977, at pp. 48f.)

These two premises led these authors to a programmatic shift in the analysis from a focus on the specification of expectations to observable firm *behavior* and the development of industries along historical trajectories (Andersen, 1994). Thus, evolutionary economics could increasingly be shaped as a 'heterodox paradigm' (Casson, 1997; Storper, 1997).

This shift in perspective from the economic (and mathematically formulated) models to a focus on observable firm behavior has epistemological consequences. Both the neo-classical hypothesis of profit maximization by the operation of the market and Schumpeter's hypothesis of the upsetting dynamics of innovations were formulated as analytical perspectives. These theories specify expectations. However, the theory of the firm focuses on observable variation. The status of the model thus changed: analytical idealizations like factor substitution and technological development cannot be expected to develop historically in their ideal-typical forms.

Nelson and Winter's first premise proposed focusing on the observables not as an *explanandum*, but as *variation* to be selected in selection environments (second premise). Innovation is then no longer to be explained, but trajectory formation among innovations serves as the *explanandum* of the first of the two 'conformable theories.' Trajectories enable enterprises to retain competences in terms of routines. Under evolutionary conditions of competition, one can expect the variation to be organized by firms along trajectories. From this perspective, however, the knowledge base is considered as completely embedded in the *institutional* contexts of the firm. (As noted above in Figure 4, I would prefer to consider this institutional environment as providing a knowledge infrastructure.) The relations between the evolutionary and the institutional perspective were thus firmly engraved in this research program (Nelson, 1994).

The supra- and inter-institutional aspects of organized knowledge production and control (e.g., within scientific communities) are considered by Nelson and Winter (1977, 1982) as part of the selection environment. However, science and technology develop and interact at a global level with a dynamics different from institutional contexts. In the Nelson and Winter model, the economic uncertainty and the technological uncertainty cannot be distinguished other than in institutional terms (e.g., market versus non-market environments). These otherwise undifferentiated selection environments generate 'uncertainty' both in the phase of market introduction and in the R&D phase. Thus, the two sources of uncertainty are not considered as a consequence of qualitatively different selection mechanisms which use different codes for the selections. The potentially different selection environments—markets, politics, and knowledge—are not specified as selective subdynamics that may interact in a non-linear dynamics.

In other words, Nelson and Winter's models are formulated in terms of the biological metaphor of variation and selection (Nelson, 1995). From this perspective, selection is expected to operate blindly. Dosi (1982, at pp. 151 ff.) added the distinction between 'technological trajectories' and 'technological regimes,' but his theory otherwise remained within the paradigm of Nelson and Winter's theory due to its focus on innovative firm behavior, that is, variation. Others have elaborated these models by using aggregates of firms, for example, in terms of sectors (e.g., Pavitt, 1984).

The models in this tradition have in common that the units of analysis (e.g., an industry) are institutionally defined and variation is organized along trajectories using a set of principles which is—for analytical reasons—kept completely separate from selection. The selection environments are not considered as differentiated (and thus at variance). Interactions among the various selection environments therefore cannot be specified. Technological innovation is considered as endogenous to firm *behavior*; trajectories are conceptualized as routines of firms. The technological component in the selection environments can consequently not be appreciated as a global effect of the networked interactions among firms, universities, and governments.

In a thorough reflection on this 'neo-Schumpterian' model, Andersen (1994) noted that firms (and their aggregates in industries) cannot be considered as the evolving units of an economy. He formulated his critique as follows:

> The limitations of Nelson and Winter's (and similar) models of evolutionary-economic processes are most clearly seen when they are confronted with the major alternative in evolutionary modeling which may be called 'evolutionary games.' […] This difference is based on different answers to the question of "What evolves?" Nelson and Winter's answer is apparently 'organisational routines in general' but a closer look reveals that only a certain kind of routines is taken into account. Their firms only interact in anonymous markets which do not suggest the playing of strategic games—even if the supply side may be quite concentrated. (Andersen, 1994, at p. 144).

While an institutional model can legitimately begin with studying observables (and is thus 'history friendly'; Malerba *et al.*, 1999, at pp. 26f.), studies about evolutionary games begin with highly stylized starting points (Andersen, 1994). These abstract assumptions can be made comparable with and traded-off against alternative hypotheses, such as the hypothesis of profit maximization prevailing in neo-classical economics. For example, one can ask to what extent an innovation trajectory can be explained in terms of the operation of market forces, in terms of its own internal dynamics of innovation, and/or in terms of interactions among the various subdynamics.

If selection mechanisms other than market choices can be specified—for example, in the case of organized knowledge production and control—the interactions among these different selection mechanisms can also be made the subject of simulation studies. In other words, the selection mechanisms span a space of possible events. The coordination mechanisms (or selection environments) cannot be observed directly, but they can be hypothesized. The model thus becomes more abstract than an institutional one. From this perspective, the observable trajectories can be considered as the historically stabilized results of selective structures operating upon one another, for example, in processes of mutual shaping. The evolutionary progression is a result of continually solving puzzles at the interfaces among the subdynamics. Thus, the routines and the trajectories can be explained from a systems-theoretical perspective as potentially special cases among other possible solutions. This next-order perspective, however, is necessarily knowledge-based because it is based on a reflexive turn (Kaufer & Carley, 1993; Scharnhorst, 1998).

## 8.2 User-producer relations in systems of innovation

In an evolutionary model one can expect mechanisms to operate along a time axis different from the one prompted by the neo-classical assumption of profit maximization prevailing at each moment of time. While profit maximization by agency remains pervasive at the systems level, this principle cannot explain the development of rigidities in the market like trajectories along the time axis (Rosenberg, 1976). In an evolutionary model, however, the (potentially stabilizing) subdynamic along the time axis has to be specified in addition to market clearing at each moment. Thus, a second selection environment over time was defined in this neo-evolutionary model.[12] (In a later section, I shall specify a third selection mechanism as operating against the axis of time.)

In his study of 'national systems of innovation' Lundvall (1988) argued that the learning process in interactions between users and producers provides a *second micro-foundation* for the economy different from the neo-classical basis of profit maximization by individual agents. He formulated as follows:

> The kind of 'microeconomics' to be presented here is quite different. While traditional microeconomics tends to focus upon decisions, made on the basis of a given amount of information, we shall focus upon a *process of learning*, permanently changing the amount and kind of information at the disposal of the actors. While standard economics tends to regard optimality in the allocation of a given set of use values as the economic problem, *par préférence*, we shall focus on the capability of an economy to produce and diffuse *use values with new characteristics*. And while standard economics takes an atomistic view of the economy, we shall focus upon the *systemic interdependence* between formally independent economic subjects. (Lundvall, 1988, at pp. 349f.)

After arguing—with a reference to Williamson's (1975, 1985) theory of transaction costs in organizations—that interactions between users and producers belonging to the same national system may work more efficiently for reasons of language and culture, Lundvall (1988, at pp. 360 ff.) proceeded by proposing the nation as the main system of reference for innovations. Optimal interactions in user-producer relations enable developers to reduce uncertainties in the market more rapidly and over longer stretches of time than in the case of less coordinated economies (Hall and Soskice, 2001; Teubal, 1979). This was discussed above when defining the function of niches.

Lundvall's theory about user-producer interactions as another micro-foundation of economic wealth production at the network level can be considered as an epistemological contribution beyond his empirical focus on national systems. The relational system of reference for the micro-foundation is different from individual agents with preferences. From this perspective, the concept of 'systems of innovation' could also be generalized to cross-sectoral innovation patterns and their institutional connections (Carlsson and Stankiewicz, 1991; Edqvist, 1997; Whitley, 2001). Because of this variety of possible relations, however, user-producer relations contribute to the creation and maintenance of a system as only one of the possible subdynamics. Other relations may also be relevant. For example, the relation between the user (in this case, the adopter) to the innovation itself—rather than to its producer—has been explored by Rogers (1962) and other researchers of the diffusion of innovations.

---

[12] The comparison among different states (e.g., using different years) can be used for comparative static analysis, but the dynamics along the time axis are then not yet specified.

In an early stage of the development of a technology, a close relation between technical specifications and market characteristics in user-producer interactions can provide a specific design with a competitive advantage (Saviotti, 1996; Rabehirosoa and Callon, 2002). In other words, proximity can be expected to serve the incubation of new technologies. However, the regions of origin do not necessarily coincide with the systems that profit from these technologies at a later stage of development. As noted above, various Italian industrial districts provide examples of this flux. As local companies develop a competitive edge, they have tended to move out of the region, generating a threat of deindustrialization. This threat has continuously to be countered by these industrial districts (Dei Ottati, 2003; Sforzi, 2003).

Analogously, this mechanism is demonstrated by the four regions designated by the EU as 'motors of innovation' in 1988. These four regions—Catalonia, Lombardia, Baden-Württemberg, and Rhône-Alpes—were no longer the main loci of innovation in the late 1990s (Krauss and Wolff, 2002; Laafia, 1999; Viale and Campodall'Orto, 2002, at pp. 162 ff.). Such observations indicate the occurrence of unintended consequences: bifurcations are generated when the diffusion dynamics of the market becomes more important than the local production dynamics. Diffusion may reach the level of the global market, and thereafter the globalized dimension can increasingly feed back on local production processes, for example, in terms of deindustrialization (Beccatini *et al.*, 2003). Given the globalization of a dominant design, firms may even compete in their capacity to destroy knowledge bases from a previous period (Frenken & Leydesdorff, 2000).

In summary, a system of innovation defined as a localized nation or a region can be analyzed in terms of user-producer relations or more aggregated in terms of institutional networks, and the stocks and flows contained in this system. However, control and the consequent possibility of appropriation of competitive edges emerge from a recombination of institutional opportunities and functional requirements. In some cases and at certain stages of the innovation process, local stabilization in a geographic area may prove beneficial, for example, because of the increased puzzle-solving capacity in a niche. However, at a subsequent stage this advantage may turn into a disadvantage because the innovations may become increasingly locked into the local conditions. As various subdynamics compete and interact, the expectation is that a more complex dynamics will emerge. Therefore, the institutional perspective on a system of innovation has to be complemented with a functional analysis. A focus on the geographical perspective of national systems without this awareness of changing boundaries can be counterproductive (Bathelt, 2003).

The 'Mode-2' thesis of the new production of scientific knowledge (Gibbons *et al.*, 1994) implies that the contemporary system has more recently gained a degree of freedom under the pressure of globalization and the new communication technologies. What seemed institutionally rigid under a previous regime (e.g., nation-states) can be made flexible under the 'globalizing' regime of communications. In a follow-up study, Nowotny *et al.* (2001, at pp. 121 ff.) specified that the new flexibility is not to be considered only as 'weak contextualization.' These authors argue that a system of innovation is a construct that is continuously undergoing reconstruction and can be reconstructed even *in the core of its operations*. This 'strong contextualization' not only affects instantaneous decisions, but also the longer-term structures in the selection processes over time (pp. 131f.). The possibilities for novelty and change are limited more in terms of our current capacity to reconstruct expectations than in terms of historical constraints.

How does one allocate the capacities for puzzle-solving and innovation across the system when the system boundaries become so fluid? The authors of the Mode-2 thesis answer as follows:

> There is no longer only one scientifically 'correct' way, if there ever was only one, especially when—as is the case, for instance, with mapping the human genome— constraints of cost-efficiency and of time limits must be taken into account. There certainly is not only one scientifically 'correct' way to discover an effective vaccine against AIDS or only one 'correct' design configuration to solve problems in a particular industry. Instead, choices emerge in the course of a project because of many different factors, scientific, economic, political and even cultural. These choices then suggest further choices in a dynamic and interactive process, opening the way for strategies of variation upon whose further development ultimately the selection through success will decide. (Nowotny *et al.*, 2001, at pp. 115f.)

The perspective, consequently, is changed from interdisciplinary—that is, based on careful translations among different discourses—to *transdisciplinary*—that is, based on an external management perspective. The global perspective provides us with more choices than were realized hitherto. This global perspective emerges from reflexive communications. Reflexive communications add another dimension to (the sum of) agency-based reflections and thus can generate a dynamics of 'reflexive' or 'radical' modernization (Beck *et al.*, 1994; Beck, Bons, and Lau, 2003, at pp. 1-3).

While Lundvall (1988) had focused on interaction and argued that communications can stabilize the local innovation environment for agents, the authors of the Mode-2 thesis argued that reflexive communications can provide us with a global perspective on the relevant environments. This global perspective enables us to assess the historically grown opportunities from the perspective of hindsight. In other words, the global perspective adds a dynamic that is different from the historical one which follows the time axis. While the latter focuses on the opportunities and constraints of a given unit (e.g., a firm or region) in its historical context, the reflexive discourse enables us to redefine the systems of reference by contextualizing and analyzing the subjects under study with hindsight. Thus, the focus shifts from the historical reconstruction of a system by 'following the actors' (Latour, 1987) to the analysis of an innovation system operating in the present. The robustness of a construct would no longer depend on its historical generation, but on the present level of support that can be mobilized in terms of expectations from the various subsystems of society (e.g., the economy or the regulatory systems involved).

Proponents of the 'Mode-2' thesis focus on this reflexive subdynamic and claim its priority over the other subdynamics including the market. The knowledge-based subdynamic of innovations is integrated with the political subdynamic of control into a 'transdisciplinary' perspective and the complex system under study is analyzed on the basis of an *a priori* priority of this reflexive perspective over social development. In other words, the complexity is reduced from the perspective of choosing the specific window of learning on the complex dynamics. The claim of this encompassing perspective among a variety of perspectives has appealed to policy makers (Hessels and Van Lente, 2008). From the Triple Helix perspective, this reduction of uncertainty by choosing a single perspective—on the basis of the assumption of a prevailing process of de-differentiation—means unnecessarily sacrificing explanatory power. What needs to be explained, are the interactions among the different perspectives.

The transformation of the political system under 'Mode-2' conditions does not entail, in my opinion, that this subdynamic can be expected to develop hierarchical control. Indeed, this might reify an analytical perspective. The 'Mode-2' thesis focuses on a reflexive turn. The political subdynamic, however, remains part of the system which is supposedly to be steered from this perspective. Because of the nested dynamics, the political subsystem can function in some instances as the steering variable, and at a next moment of time encounter itself as a dependent variable; for example, in the case of unintended consequences. In the complex dynamics of communication, unintended consequences of specifically coded communications can be expected to prevail because each communication is part of a distribution of communications necessarily containing uncertainty. Since the complex dynamics consists of subdynamics which select upon each other, the system can also be resilient against steering, and in some phases more than in other (Van den Daele *et al.*, 1979; Weingart, 1997). Thus, the question of a empirical study (modeling and simulation) of systems of innovation remains crucial.

Not incidentally, the optimism of the 'Mode-2' thesis—which did not imply the need to do systematic research beyond the telling of success stories ('best practices')—resonated with aspirations at the level of the European Commission. Confronted with stubborn national rigidities especially in the arena of science policy, a theory which legitimated a perspective of 'trans-nationality' and 'transdisciplinarity' could count on a warm welcome. Because of the subsidiary principle in the EU which specifies that the Union should leave to national governments what does not require harmonization at the transnational levels, science policy initiatives had successfully been defended by national scientific elites against European intervention during the 1980s (Mulkay, 1976). The European Committee had circumvented this blockage by focusing on *innovation* policies—as different from science and technology policies—in the successive

Framework Programs. In addition to demanding that two or more nationalities be represented in the bids for these programs, collaborations between universities, industries, and public research centers were required before one could qualify for obtaining a grant. From the perspective of 'pure science,' this focus on 'development' further legitimated a divide between the national systems of research councils—which are controlled by national scientific elites—and the grey and allegedly 'bureaucratic' procedures of the European Commission.

Within this context, the 'Mode-2' thesis provided further legitimacy to the aspirations of the European Science Foundation and a European system of research councils which would balance US systems such as the National Science Foundation and the National Institute of Health. In the USA, however, these systems are both federal and national, and therefore traditionally controlled by scientific elites at the national level (Brockman, 1995). In the European configuration, a trans-national system of quality control would also require the construction of a trans-disciplinary frame of reference. The bibliometric framework of citation analysis cannot provide this frame of reference, because this would reproduce traditional differences among disciplines and national cultures in Europe. For example, the more internationally oriented national countries such as in Scandinavia, the UK, Ireland, and the Netherlands could then dominate the framework.

In this context, the 'Mode-2' thesis provided a promising perspective: all systems can be deconstructed as specific constructs using reflexive policy analysis, and then the margins for deliberate intervention in the reconstruction would depend on the decomposing power of the reflexive discourse. However, this underestimates the problems involved in proceeding from discursive reconstruction to deliberate action. The latter presumes informed choices among and decisions about ranges of options that reproduce the complexity under study according to a complex dynamics that is different from its politically controlled subdynamic.

What did the 'Mode-2' model add to the model of 'national innovation systems' in terms of providing another micro-foundation? Lundvall's micro-economics were grounded in terms of interactions between users and producers rather than in terms of the individual preferences of agents. The authors of the 'Mode-2' thesis defined another communication dynamic relevant to the systems of innovation. This other perspective is possible because a network contains a dynamic both at the level of the nodes and at the level of the links. While agency can be considered as a source or recipient of communication—and can be expected to be reflexive, for example, in terms of learning and entertaining preferences—an agent has a contingent position at a node in the network (Burt, 1982). The links of a communication system, however, operate differently from the nodes in the network. Links can be replaced for functional reasons and densities in the networks can thus migrate as an unintended consequence.

Concepts like reflexivity and knowledge have different meanings from one layer of the network to another and these different layers can be made the subject of other discourses. For example, agents entertain preferences, but the structure of the network of communications provides some agents with more access than others. In addition to actions which generate the variations, the dynamics of communications, that is, at the level of the links, are transformative (Bhaskar, 1979; 1998, at p. 207), and therefore change (and may innovate!) the structural mechanisms of selection and coordination. These changes are endogenous to the network because they can be the result of non-linear interactions among previously stabilized aggregates of actions. Recursions and interactions add non-linear terms to the results of micro-actions.

Luhmann (1984) was the first to propose that communication *among* agents be considered as a system of reference different from agency. More specifically, one can expect the dynamics of inter-human coordination by communication to be different from the reflexive consciousness of agents' perceptions (e.g., Luhmann, 1996). Using a terminology of Maturana and Varela (1980; 1984), consciousness at the level of agency and communication can be considered as 'structurally coupled' in the events. An interaction can be attributed as an action to an actor, while it can be expected to function as a communication within a communication system (Maturana, 1978; Leydesdorff, 2001). However, the reflection by a social system operates differently from reflection at the level of individual consciousness. Human language enables us not only to provide meaning to the communication of information, but also to communicate meaning on top of the first-order dynamics of information.

Each coordination mechanism provides its own meaning to information by invoking its specific code of communication (Malone & Crowston, 1994). Words may have different meanings in other contexts. For example, in scientific communications 'energy' has a meaning different from its meaning in political discourse. While economists and politicians may worry about 'shortages of energy,' 'energy' is defined as a conserved quantity in physics. The evolutionary dynamics of social communication can add another layer of complexity to the first-order dynamics of information exchanges among agents.

The self-organization of communication into various (functionally different) coordination mechanisms on top of the institutional organization of society in a national system enables the social system to process more complexity than in an organizationally controlled mode. However, under this condition one can expect to lose increasingly the notion of accountable centers of coordination; central coordination is replaced with a number of more abstract and interacting coordination mechanisms. The interacting (sub)systems of communication can become increasingly differentiated in terms of their potential functions for the self-organization of the social system. This communication regime reshapes the existing communication structures as in a cultural evolution. In other words, selection mechanisms other than 'natural' ones begin to reconstruct the system from the various perspectives of the respective coordination mechanisms.

In summary, the communicative layer provides society with a set of selection environments for historical institutions. In the case of communication systems, however, selections operate probabilistically, that is, with uncertainty. Translations among the differently coded communication may reduce the uncertainty. Thus, these selection mechanisms can only be specified as hypotheses. The specification of these expectations, however, guides the observations in terms of specifiable uncertainties (e.g., Leydesdorff

& Fritsch, 2006). Furthermore, these communication dynamics of the social system are complex because the codes of the communication have been differentiated historically (Strydom, 1999).

Communications develop along the functionally different axes, but they can additionally be translated into each other by using the different codes at interfaces reflexively. Thus, systems of translation are generated. For example, interaction terms among codes of communication emerged as a matter of concern within knowledge-based corporations when interfaces between R&D and marketing had increasingly to be managed (Galbraith, 1967). In university-industry-government relations three types of communications are interfaced. Let me now turn to my thesis that the utilization of the degrees of freedom between institutions and functions among the three subsystems interacting in a Triple Helix enables us to understand these processes of innovation.

*8.4 A Triple Helix model of innovations*

The systems-of-innovation approach defined innovation systems in terms of institutional units of analysis. 'Mode-2' analysis defined innovations exclusively in terms of reconstructions on the basis of emerging perspectives in communication. The Triple Helix approach combines these two perspectives as different subdynamics of the systems under study. However, this model enables us to include the dynamics of the market as a third perspective. As noted, the perspective of neo-classical economics is micro-founded in the natural preferences of agents. Thus, one can assume that innovation systems are driven by various subdynamics to varying extents. Consequently, the discussion shifts from a philosophical one about what an innovation system 'is,' or the question of how it should be defined, to the methodological question of how one can study innovation systems in terms of their different dimensions and subdynamics.

Within this complex dynamic, the two mechanisms specified above—user-producer interactions and reflexive communications—can be considered as complementary to the micro-foundation of neo-classical economics. First, each agent or aggregate of agencies is positioned differently in terms of preferences and other attributes. Second, the agents interact, for example in economic exchange relations. This generates the network perspective. Third, the arrangements of positions (nodes) and relations (links) can be expected to contain information because not all network positions are held equally and links are selectively generated and maintained. The expected information content of the distributions can be *recognized* by relevant agents at local nodes. These recognitions provide meaning to the events and these meanings can also be communicated. The recognition thus generates knowledge bases both at the addresses of the agents, in their organizations, and at the level of society. Knowledge can also be processed as discursive knowledge in the network of exchange relations. Figure 6 summarizes this configuration.

With this visualization I intend to make my argument epistemologically consistent by relating the above reflections to the underlying dimensions of the Triple Helix model. The three analytically independent dimensions of an innovation system were first distinguished in Figure 3 (above) as (1) the geography which organizes the positions of agents and their aggregates; (2) the economy which organizes their exchange relations; and (3) the knowledge content which emerges with reference to either of these dimensions (Archer, 1995). Given these specifications, we were able to add the relevant interaction terms. The second-order interaction among these interactions then provides us with the hypothesis of the development of a knowledge base endogenous to the system under study.

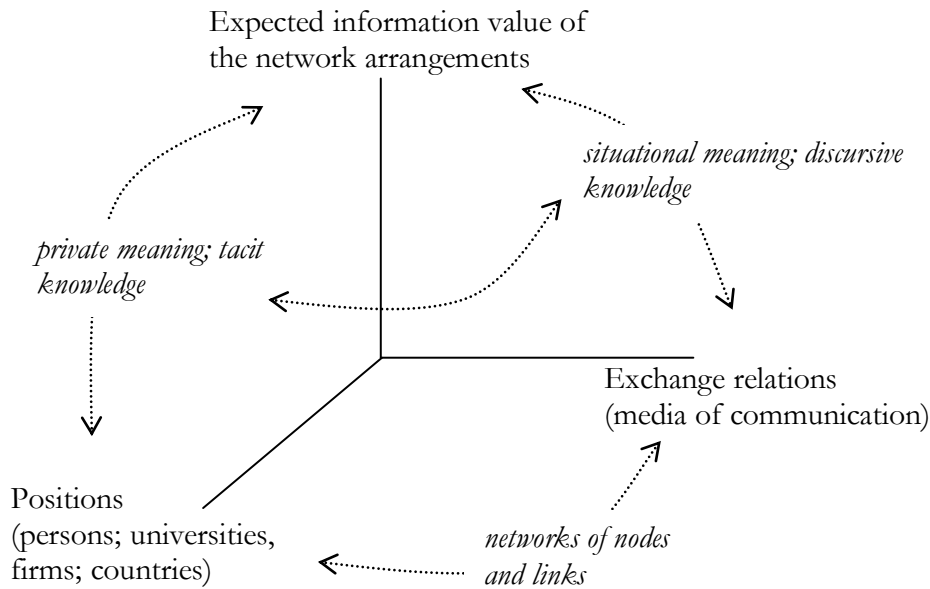

**Figure 6:** Micro-foundation of the Triple Helix Model of Innovations

Figure 6 specifies the knowledge base as an interaction between discursive and tacit knowledge. Along the three axes, the three micro-operations are represented. Each agency (agents, institutions, nations) has a position in the network and can be considered as naturally gifted with a set of preferences. This assumption accords with the micro-foundation of neo-classical economics. The network dynamics are first micro-founded in terms of natural preferences. Learning in relations, however, can change both the agents and their institutions by embedding them in specific (e.g., non-market) contexts. The second (horizontal) axis thus corresponds with Lundvall's micro-foundation in user-producer interactions. National systems of innovation can then be considered as specific forms of organization which reduce transaction costs (Williamson, 1975, 1985). For example, the Scandinavian environment might generate an institutional framework which changes transaction costs to such an extent that a second independent dynamics can be sustained in addition to the market mechanism.

'Mode-2' can be considered as the third axis: the interacting dynamics can be made the subject of reflexive analysis. I drew this third axis as vertical in Figure 6 because the emergence of a *meta*-perspective is the implicit assumption of the 'Mode-2' thesis. However, one can rotate the figure and change the order of the axes without any loss of explanatory power: three micro-mechanisms are involved: one based on positions, a second based on the possibility of exchange, and a third based on the possibility for agent-based and discursive learning.

When the learning is grounded in agency, (social) psychological mechanisms and categories can be used for the analysis. However, when the learning is carried by distributions in networks, the socio-psychological categories provide us only with metaphors, but the operations have to be specified differently because networks can be expected to contain a dynamics different from agency. For example, agency tends to integrate conflicting perspectives by making trade-offs. Networks allow for other solutions, such as differentiation when different control mechanisms are available. For example, normative and analytical considerations can be entertained, distinguished, and traded-off at different positions in the network. The second-order interaction between learning in individuals and networks generate configurational knowledge as a next-order regime of expectations.

## 9. Empirical studies and simulations using the TH model

Unlike biological models that focus on variation with reference to 'natural' selection mechanisms, the Triple Helix model focuses primarily on the specification of different selection mechanisms. Selection is structural. Three helices are sufficiently complex to understand the social reproduction of the dynamics of innovation (Leydesdorff, 2009; cf. Lewontin, 2000). What is observable can be specified as relative equilibria at interfaces between two selection mechanisms operating upon each other. When repeated over time, each co-variation can be developed into a co-evolution, and a next-order, that is, more complex, system can be generated in a process of mutual shaping among the interactions.

I have argued that the Triple Helix can be elaborated into a neo-evolutionary model which enables us to recombine sociological notions of meaning processing, economic theorizing about exchange relations, and insights from science and technology studies regarding the organization and control of knowledge production. The further codification of meaning in scientific knowledge production can add value to the economic exchange relations (Foray, 2004; Frenken, 2005). The model can serve as a heuristics. Its abstract and analytical character enables us to explain current transitions towards a knowledge-based economy as a new regime of operations. The Triple Helix model thus substantiates and operationalizes the general notion of a knowledge-based economy as a self-organizing system (Krugman, 1996).

The differentiation in terms of selection mechanisms can be both horizontal and vertical. Vertically the fluxes of communications are constrained by the institutional arrangements that are shaped in terms of stabilizations of previous communications. Horizontally, the coordination mechanisms can be of a different nature because they can be expected to use different codes. For example, market transactions are different from scientific communications. Market transactions can also be cross-tabulated with organizational hierarchies (Williamson, 1985; Lundvall, 1988). While the control mechanisms at interfaces can be considered as functional for the differentiation among communications, the hierarchy in the organization may help to reduce the problem of coordination between functions to a multi-level problem within the institutional dimension.

In summary, the functional perspective is different from the institutional one. Functional communications evolve; institutional relations function as retention mechanisms which respond to functional incentives. The specification of functions in the socio-economic analysis requires reflexivity. All reflections can again be made the subject of communication. Thus, one can study a Triple Helix at different levels and from different perspectives. For example, one can study university-industry-government relations from a (neo-)institutional perspective (e.g., De Rosa Pires and De Castro, 1997; Etzkowitz *et al.*, 2000; Gunasekara, 2006) or one can focus on the relations between university science and the economy in terms of communications (e.g., Langford *et al.*, 1997). Different interpretations of the Triple Helix model can be at odds with each other and nevertheless inform the model. Each metaphor stabilizes a geometrical representation of an otherwise more complex dynamics.

Competing hypotheses derived from different versions of the Triple Helix can be explored through formal modeling and appreciated through institutional analysis. The case studies inform the modeling efforts about contingencies and boundary conditions, while simulation models enable us to relate the various perspectives. Such translations potentially reinforce the research process by raising new questions, for example, by comparing across different contexts and/or with reference to emerging phenomena. From this perspective, innovation can be considered as the reflexive recombination at an interface, such as between a technological option and a market perspective. Specification of the different contexts, however, requires theorizing. For the purpose of innovation, the perspectives have to be combined, for example, in terms of a plan.

The three strands of the Triple Helix are treated as formally equivalent in the model, but they are substantially very different. The selection mechanisms are expected to operate asymmetrically. The one strand (university) is institutionally less powerful than the other two strands. Furthermore, the other two strands (government and industry) are increasingly and indirectly co-opting the university in a variety of ways, even if one disregards the direct influence of the so-called military industrial complex. However, the university has specific strengths: first, it is salient in providing the other two systems with a continuous influx of new discursive knowledge (e.g., papers and patents) and new knowledge carriers (students). From this perspective, the university can be considered as the main carrier of the knowledge-based innovation system (Godin and Gingras, 2000). Knowledge-based fluxes continuously upset and reform the dynamic equilibria sought by the two other strands of the political economy.

The Triple Helix model is sufficiently complex to encompass the different perspectives of participant observers (e.g., case histories) and to guide us heuristically in searching for options newly emerging from the interactions. What is the contribution of this model in terms of providing heuristics to empirical research? First, the neo-institutional model of arrangements among different stakeholders can be used in case study analysis. Given the new mode of knowledge production, case studies can be enriched by addressing the relevance of the *three* major dimensions of the model. This does not mean to disclaim the legitimacy of studying, for example, bi-lateral academic-industry relations or government-university policies, but one can expect more interesting results by observing the interactions among the three subdynamics. Secondly, the model can be informed by the increasing understanding of complex dynamics and simulation studies from evolutionary economics (e.g., Malerba *et al.*, 1999; Windrum, 1999). Thirdly, the Triple Helix model adds to the meta-biological models of evolutionary economics the sociological notion of meaning being exchanged among the institutional agents (Habermas, 1987; Leydesdorff, 2001; Luhmann, 1984 [1995a]).

Finally, on the normative side of developing options for innovation policies, the Triple Helix model provides us with an incentive to search for *mismatches* between the institutional dimensions in the arrangements and the social functions carried by these arrangements. The frictions between the two layers (knowledge-based expectations and institutional interests), and among the three domains (economy, science, and policy) provide a wealth of opportunities for puzzle solving and innovation. The evolutionary regimes are expected to remain in transition because they are shaped along historical trajectories. A knowledge-based regime continuously upsets the political economy and the market equilibria as different subdynamics. Conflicts of interest can be deconstructed and reconstructed, first analytically and then perhaps also in practice in the search for solutions to problems of economic productivity, wealth retention, and knowledge growth.

The rich semantics of partially conflicting models reinforces a focus on solving puzzles among differently codified communications reflexively. The lock-ins and the bifurcations are systemic, that is, largely beyond control; further developments are based on the self-organization of the interactions among the subdynamics. The subdynamics can also be considered as different sources of variance which disturb and select from one another Resonances among selections can shape trajectories in co-evolutions, and the latter may recursively drive the system into new regimes. This neo-evolutionary framework assumes that the processes of both integration and differentiation remain under reconstruction.

## 10. The KBE and the measurement

From the perspective of the information sciences, the above discussion of innovation theory and theories of technological change needs to be complemented with a further specification about the operationalization and the measurement. Can a measurement theory for the communication of meaning and knowledge in a Triple Helix model also be specified? How does the communication of knowledge differ from the communication of information and meaning, and how can these differences be operationalized? How do the communication of information, meaning, and knowledge as layers in communication systems relate and potentially operate upon one another? How does this vertical differentiation in the codification relate to the horizontal differentiation among the three (or more) coordination mechanisms in a Triple Helix model?

*10.1 The communication of meaning and information*

The idea that human beings not only provide meaning to events, but are able to communicate meaning in addition to the communication of information, emerged gradually during the 20[th] century with the development of sociology as a discipline. According to Weber (e.g., 1904, 1917) values were to be considered as the crucial domain of human encounter and social development. As is well known, Weber advocated adopting 'value freeness' as a methodological principle in the sociological analysis, while paying proper attention to the value-ladenness of the subject matter in sociological analysis (Watkins, 1952). From Weber's perspective, values govern human history as givens (Weber, 1919).

Durkheim (1912) noted in this same period that values can also be considered as 'collective consciousness.' Parsons (1968) emphasized that this concept of *another dynamic at the supra-individual level* can with hindsight be considered as constitutive for the new science of sociology. He traced it—that is, the idea that *social interaction* bestows events with qualitatively different meanings—back to American pragmatism (Mead, 1934), on the one hand, and on the other to Freud's (1911) and Durkheim's (1912) independent discoveries of the 'reality principle' and 'collective consciousness,' respectively.[13] This new sociological program of research clashed with positivism— which also finds its origins in sociology (e.g., Auguste Comte), in opposition, however, to the idealistic philosophies of the 19[th] century—because the focus was no longer on empirical data, but rather on what the data means, and how the subjects under study can sometimes reach consensus or otherwise dwell in conflicts about such meaning. The ensuing 'Positivismusstreit' in German sociology had its origins in the 1930s, but was exported to the United States by German emigrants in the prewar period (Adorno *et al.*, 1969).

In his 1971 debates with Habermas (who as a neo-marxist sided with the anti-positivists in the 'Positivismusstreit'), Luhmann (1971) proposed that the communication of meaning be considered as the core subject of sociology: coordination among human beings is not brought about by information transfer, but rather by the communication of meaning (Habermas & Luhmann, 1971). Unlike information, meaning cannot be transferred over a cable, but it can be communicated in interactions among reflexive agents. According to Luhmann (1984), sociologists should focus on the dynamics of meaning in communication (Luhmann, 2002). Habermas (1981, 1987), however, wished to focus on 'communicative action' and communicative competence as attributes of human beings.

In these exchanges, both Habermas and Luhmann made references to Husserl's reflections on 'intersubjectivity' as a common base, but they provided Husserl's philosophy with another interpretation (Husserl, 1929, 1962; Derrida, 1974). Habermas (1981, at pp. 178f.) followed Schutz (1952, at p. 105) in arguing that Husserl had failed to ground his concept of 'intersubjectivity' in interhuman communication (cf. Luhmann, 1995b, at p. 170). This grounding would require the concept of a 'life-world' in which communication is embedded. In my opinion, Luhmann remained closer to Husserl's so-called transcendental phenomenology by considering social relations as instantiations which are embedded in 'virtual,' yet structured communication fluxes. The communication dynamics explain the social relations instead of *vice versa* analyzing communication structures as a consequence of social relations in a 'lifeworld' (Schutz, 1975, at p. 72).

---

[13] According to Parsons's (1952) reading of Freud, the social environment is internalized at the level of the super-ego.

The approach which considers communications not as attribute to organizations and agency, but organizations and agency as constructed in and by interhuman communications, finds its philosophical origins in the *Cartesian Mediations,* which Husserl (1973) wrote in 1929. Husserl followed Descartes by questioning not only what it means to be 'human,' but also the referent of human intentionality. While the first question refers back to Descartes' (1637) '*cogito ergo sum,*' the latter addresses the subject of doubt, that is, the *cogitatum*: the external referent of one's doubting. For Descartes this *cogitatum* could be distinguished only negatively from the *cogito* as that which transcends the contingency of one's *cogito*. From this perspective, the other in the act of doubting is defined as God. God transcends the contingency of the *cogito*, and therefore one can expect this Other to be eternal.

Husserl proposed to consider the *cogitatum* no longer as a personal God, but as the intentional substance *among* human beings which provides the *cogito* with an horizon of meanings. We—as *cogitantes*—are uncertain about what things mean, and the communication of this uncertainty generates an intersubjectivity which transcends our individual subjectivities. Although meanings are structured at the supra-individual level, these structures are no longer identified with a personal God. On the contrary, meaning can be constructed, enriched, and reproduced among human beings by using language.[14]

By using language one is able to relate meanings to one another. However, within language the world is resurrected as an architecture in which the words can be provided with meaning at the supra-individual level. However, this meaning is not provided by the words or their concatenations in sentences or networks of co-occurrences. Language organizes the concepts by providing specific meaning to the words at specific instances (e.g., in sentences). The instantiations refer to what could have been differently constructed and understood. In other words, the *cogitata* are not specific; they remain uncertain.

Husserl emphasized that this substance of the social system ('intersubjective intentionality') is different from subjective intentionality because one knows it *ex ante* as beyond the domain of the individual. The study of this new domain—Husserl used the Leibniz's word 'monade'—might provide us with 'a concrete ontology and a theory of science' (*ibid*., at p. 159). However, Husserl conceded that he had no instruments beyond

---

[14] Husserl acknowledged this function of language in the generation of meaning when he formulated for example: 'The beginning is the pure and one might say still mute experience which first has to be brought into the articulation of its meaning' (*ibid*., p. 40).

the transcendental apperception of this domain and therefore he had to refrain from empirical investigations:

> We must forgo a more precise investigation of the layer of meaning which provides the human world and culture, as such, with a specific meaning and therewith provides this world with specifically 'mental' predicates. (Husserl, 1929, at p. 138; my translation).

In my opinion, two important developments in applied mathematics have made it possible to address the questions which Husserl felt as beyond his reach: first, Shannon's mathematical theory of communication provided us with categories for analyzing communications in terms of uncertainties (Abramson, 1963; Theil, 1972; Leydesdorff, 1995), and, second, Rosen's (1985) mathematical theory of anticipatory system and Dubois's (1998) elaboration of this theory into the computation of anticipatory systems provided us with categories for studying the evolution of systems which are based on expectations and their potential functions for further developing codified communications (Leydesdorff, 2008, 2009).

## 10.2 The expectation of social structure

In addition to these methodological advances, some theoretical steps in sociology were crucial. First, Parsons (1951, at p. 94; 1968, at p. 436) elaborated the communication of meaning in terms of what he called a 'double contingency' in interhuman relations. *Ego* relates to *Alter* not only in terms of observable relations, but also in terms of expectations. *Ego* expects *Alter* to entertain expectations like those *Ego* finds in her own mind. The two systems (*Ego* and *Alter*) expect each other to operate in terms of expectations. While *Ego* and *Alter* are defined at the level of individual consciousness, Luhmann (1984, 1986) generalized this model as the model of communication between and among meaning-processing systems (Vanderstraeten, 2002). From this perspective, the Triple Helix model can also be considered as representing a triple contingency among three communication systems (Strydom, 1999; Leydesdorff, 2008).

In addition to the model of double contingency, Luhmann used Parsons's (1963a, 1963b, 1968) further elaboration of mediation in terms of symbolically generalized codes of communication. For example, money enables us to make economic transactions without having to discuss the price. Using Maturana & Varela's (1980, 1984) theory of *autopoiesis,* Luhmann (1984 [1995a], 1990, and 1997) elaborated a sociological theory of the dynamics of codified communications (Leydesdorff, 2001). Crucial is that meaning can be communicated among human beings, and that the coordination of this communication can become self-organizing—that is, beyond the control of the communicating agents—under the condition of modernity, that is, under the pressure of the functional differentiation of the codes of communication in the various coordination mechanisms.

The differentiation in the codification (e.g., among economic exchange relations, political communication, and scientific communication) generates a feedback that changes the organization and dynamics of the social system. However, Luhmann (1990, at p. 340) added that 'developing this perspective is only possible if an accordingly complex systems-theoretical arrangement is specified.' In my opinion, this requires an information-theoretical elaboration of this sociological theory of communication. Information theory provides us with a mathematical apparatus to study communication systems at a more abstract level both in terms of the composition and in terms of dynamic developments (Bar-Hillel, 1957; Theil, 1972; Brooks & Wiley, 1986; Leydesdorff, 1995; Yeung, 2008). One can distinguish layers and dimensions at different moments of time, for example, by using subscripts and superscripts as indices.

Structure in the data provides meaning. Structure can be analyzed in observed data by using multivariate statistics, for example, factor analysis. The factor analysis reduces the data by focusing on the latent dimensions ('eigenvectors') of the networks of relations among observable vectors. The eigenvectors, in other words, provide us with a second-order dynamics in terms of which changes at the level of relations among vectors—that is, the first-order dynamics in terms of observable data—can be provided with meaning.

Note that meaning can be provided both by an analyst and by a participant-observer within the system under study. In both cases, the relational data is positioned in a vector space with a topology very different from the relational space in which the network is constructed. For example, the vector space is based on continuous coordinates, while graph-theoretical network analysis is based on discrete events. The different topology of the vector space enables us to formalize the concepts of meaning and meaning-processing. This constructed space can be considered as a *cogitatum*. It remains a construct that cannot be observed directly. The function of this constructed space is to enable us to communicate intersubjectively the meaning of what one is able to observe in the first contingency of a relational space. Social structure is thus created as a cultural order of expectations.

Changes in the relational space communicate Shannon-type information. Meaning is communicated within a vector space. For example, proximity in terms of positions may make a difference in terms of how easily meaning can be communicated across relations (Freeman, 1978/1979; Burt, 1995). The distinction between a first- and a second-order dynamics of communicating information and meaning, respectively, enables us also to specify the mutual information between these two dynamics as meaningful information in the sense of Bateson's (1972, at p. 453) dictum of 'a difference which makes a difference.'

A difference in the data can make a difference in terms of the organization of the data, and thus be meaningful. In other words, the organization of the data provides the (Shannon-type) information with meaning. However, not all uncertainty is meaningful, and thus a selection is involved. Selection is structural, but operates as a *cogitatum*. The Darwinian notion of 'natural' selection is replaced with the hypothesis of different selection environments. Neither the selection environments as structures nor their dynamic functions should be reified from this perspective; they remain constructs and orders of expectations. The function of specifying these selection environments is to enable theoretical discourse to enrich the development by providing it with relevant distinctions in the knowledge base. For example, the distinction between market *versus* non-market selection environments induced the emergence of evolutionary economics as a theoretical discourse in the 1980s (Nelson & Winter, 1977, 1982; Pavitt, 1984).

## 10.3 Configurations in a knowledge-based economy

This operationalization of the communication of meaning in a space with different characteristics provides us with a next step in the full specification of the Triple Helix model. Meaning-providing subsystems can be considered as carried by the eigenvectors of the networks, that is, as densities in the structures of communications. In the static model the eigenvectors can be spanned orthogonally. Dynamically, they represent selective structures. The observable variation at the first-order level provides the co-variation among these selective structures. For example, in Figure 1 above, patents were positioned as observable events in a vector space spanned by the three dimensions of the Triple Helix.

When three or more dimensions can operate upon one another, mutual information or co-variation in more than two dimensions can also be computed. The resulting information or more-dimensional co-variance can be either positive or negative (McGill & Quastler, 1955; Yeung, 2008). A negative expected information value of this measure would reduce uncertainty and can therefore be considered as configurational information (McGill, 1954). Configurational information provides us with an indicator of the interaction among different (and potentially orthogonal) dimensions.[15] Since this reduction of uncertainty can be attributed to the developments in and interactions among the (sub)systems under study, it can be considered as an indicator of self-organization, that is, the local production of negative entropy.

Note that this reduction of uncertainty at the level of a configuration among functions is analytically different from reduction of uncertainty provided by the factor analysis. The factor analysis reduces the data by capturing the common variances among the variables. This first-order reduction of uncertainty can be considered as a structure in the data to which one can ascribe a semantic meaning (for example, by designating the factors). The reduction of uncertainty because of the configuration among the main dimensions (eigenvectors of the data matrix) distinguishes in a next-order process among meanings which make a difference, and thus indicates the extent to which knowledge as codified meaning can be expected to operate within a system.

---

[15] Reduction of uncertainty at the systems level contradicts the Second Law which holds equally for probabilistic entropy $H$ (Theil, 1972: 59 ff.). According to Gibbs entropy formula $S = k_B H$. The Boltzmann constant $k_B$ is multiplied with $H = - \Sigma_i p_i \log p_i$. Multiplication by the Boltmann constant provides the otherwise dimensionless Shannon-entropy $H$ (expressed in bits) with the dimension of Joule/Kelvin. However, the dynamics of the Second Law are not caused by the Boltzmann constant for the very reason that this is a constant which remains external to the development of the probabilistic entropy $H$.

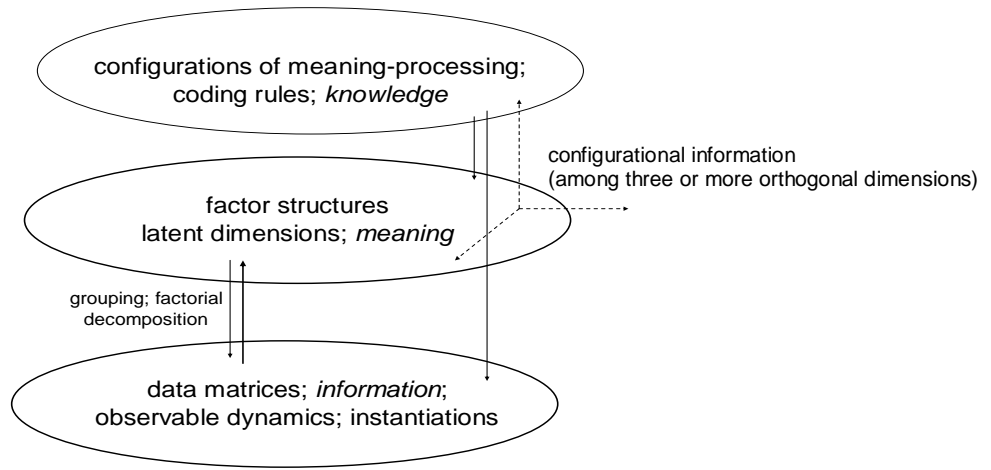

**Figure 7**: A layered process of codification of information by meaning, and codification of meaning in terms of a knowledge base.

Figure 7 summarizes the model. This model includes both horizontal and vertical differentiation. The horizontal differentiation among latent coordination mechanisms was considered by Luhmann (1997) as functional differentiation and is operationalized as the orthogonal dimensions which result from the factor analysis. Vertically, the model distinguishes between interactions at the bottom level, potential self-organization of the communications in different configurations at the top level, and structuration by organization at the level in between.

In summary, the evolving networks of relations among agents can be considered as the retention mechanisms of flows of communication through these networks. The flows are structured by functionally different codes of communication (Luhmann, 1984 [1995a]), and enabled and constrained by structures in the data at each moment of time (Giddens, 1984). For example, a network of scientific communications may carry a flow of publications and/or patents. In a network of political communications, power and legitimacy provide the communications with differently codified meanings.

The functions of flows of communication develop evolutionarily in terms of the eigenvectors of the networks, while the networks of relations develop historically in terms of (aggregates of) actions. The functions can be operationalized as the latent dimensions (eigenvectors) of the networks of relations among the agents. However, the eigenvectors develop in a vector space with a topology and dynamics different from those of the relational space among the agents. For example, relations develop within a space, whereas the vector space can develop as a space, for example, by adding new dimensions.

The relations (and the nodes) are positioned in a vector space (Burt, 1982); positions reflect the meanings of relations (events) in the various dimensions. Reflexive agents are not only embedded in the relational space which they span, but able to provide meaning to the positions and relations. This can be considered first-order reflexivity. When this reflexivity is made the subject of theoretical reflections, next-order reflexivity is added as an overlay to this dually-layered system of relations and positions. The increased availability of this more abstract (since codified) type of communication changes the systems in which it emerges and on which it rests as another selection environment. The processes of reflexive change in such systems are enhanced by adding a knowledge base to the communication. This so-called 'radicalized' or 'reflexive' modernization (Beck *et al.*, 1994 and 2003; Giddens, 1990) can be expected to operate in some configurations more than in others. However, it tends to remain beyond the control of agents interacting in terms of relations (that is, generating uncertainty) and positions (that is, providing meaning to uncertainty).

The knowledge-base of an economy can be considered as this evolving *configuration* among the functions of coordination mechanisms. Using the Triple Helix model of university-industry-government relations, the carrying functions of a knowledge-based economy were specified above as (1) economic wealth generation, (2) knowledge-based novelty production, and (3) normative (e.g., political) control (Etzkowitz & Leydesdorff, 2000). Over time, the knowledge base thus generated—indicated as the reduction of uncertainty contained in a configuration—can be stabilized, meta-stabilized, or globalized.

Configurations can thus be distinguished in terms of the extent to which a synergy is self-organized among the main subdynamics of a knowledge-based economy. Note that with the opposite sign, configurations may also frustrate the further development of a knowledge base at the systems level by generating more uncertainty than can be absorbed by the relevant subsystems in their current configuration. This empirical research program remains necessarily a piecemeal enterprise (e.g., Leydesdorff & Sun, forthcoming).

The information sciences are crucially positioned in a configuration at a crossroad among the other relevant sciences such as economics, policy analysis, and innovation studies because of their emphasis on operationalization and measurement. The envisaged research program entails the further specification of these complex (since nonlinear) relations among the processing of uncertainty, expectations/intentions/meaning, and knowledge in communication systems, so that the knowledge base of an economy can further be measured, modeled, and explained.

## Acknowledgments
I am grateful to the comments of Diana Lucio-Arias, Andrea Scharnhorst, Wilfred Dolfsma, and a number of anonymous referees on previous drafts of this paper. The paper is partially based on a rewrite of various chapters of Leydesdorff (2006).